\documentclass[aps,prb,floatfix,amsmath,amssymb,superscriptaddress,twocolumn,nofootinbib]{revtex4-2}
\usepackage{bm}                     
\usepackage{appendix}
\usepackage[latin1]{inputenc}
\usepackage{dcolumn}      
\usepackage{bbm}
\usepackage{color}
\usepackage{xcolor}

\usepackage[mathscr]{eucal}
\usepackage{latexsym}
\usepackage{amsbsy}
\usepackage{float}
\usepackage{graphicx}
\usepackage{epsfig}
\usepackage{subfigure}
\usepackage{wrapfig}
\usepackage{epsf}
\usepackage{float}
\usepackage[normalem]{ulem}
\usepackage{tikz}
\usetikzlibrary{automata, positioning, arrows}
\usepackage{qcircuit}

\definecolor{darkblue}{HTML}{004D6B}
\definecolor{darkred}{HTML}{8c1515}
\definecolor{darkgreen}{HTML}{006400}
\usepackage{hyperref}
\hypersetup{ colorlinks=true, urlcolor=darkblue, citecolor=darkred,
    linkcolor=darkblue, breaklinks }

\newcommand{\ba}{\begin{array}}
\newcommand{\ea}{\end{array}}
\newcommand{\be}{\begin{equation}}
\newcommand{\ee}{\end{equation}}
\newcommand{\bea}{\begin{eqnarray}}
\newcommand{\eea}{\end{eqnarray}}

\begin{document}

\title{ Engineering unsteerable quantum states with active feedback }

\author{Samuel Morales}
 \affiliation{Institut f\"ur Theoretische Physik,
Heinrich-Heine-Universit\"at, 40225  D\"usseldorf, Germany}
\author{Yuval Gefen}
\affiliation{Department of Condensed Matter Physics, Weizmann Institute, 7610001 Rehovot, Israel}

\author{Igor Gornyi}
\affiliation{\mbox{Institute for Quantum Materials and Technologies, Karlsruhe Institute of Technology, 76021 Karlsruhe, Germany} }
\affiliation{\mbox{Institut f\"ur Theorie der Kondensierten Materie, Karlsruhe Institute of Technology, 76128 Karlsruhe, Germany}}

\author{Alex Zazunov} 
\affiliation{Institut f\"ur Theoretische Physik,
Heinrich-Heine-Universit\"at, 40225  D\"usseldorf, Germany}

\author{Reinhold Egger}
\affiliation{Institut f\"ur Theoretische Physik,
Heinrich-Heine-Universit\"at, 40225  D\"usseldorf, Germany}

\begin{abstract}
We propose active steering protocols for quantum state preparation in quantum circuits where each 
system qubit is connected to a single detector qubit, employing 
a simple coupling selected from a small set of steering operators. The decision is made such that the expected 
cost-function gain in one time step is maximized. We apply these protocols to several many-qubit models. 
Our results are underlined by three remarkable insights. 
First, we show that the standard fidelity does not give a useful cost
function; instead, successful steering is achieved by including local fidelity terms. Second, although the steering  
dynamics acts on each system qubit separately, entanglement in the generated target state is introduced, and can 
be tuned at will, by performing Bell measurements on detector qubit pairs after every time step.
This implements a weak-measurement variant of entanglement swapping.
Third, numerical simulations suggest that the active steering protocol can reach arbitrarily designated 
target states, including passively unsteerable states such as the $N$-qubit W state.  
\end{abstract}
\date{\today}

\maketitle
\section{Introduction}\label{sec1}

Solving complex tasks by means of control and feedback circuits is of ubiquitous 
importance in the modern world.  A famous example is the Apollo space mission where optimal control enabled a spaceship to smoothly land on the moon with vanishing target velocity.  
Can one design a similar strategy where one ``steers'' a \emph{quantum} system at will towards a
predesignated target state?  In the quantum case, exerting ``control'' over the system requires the ability to perform quantum measurements while
 ``feedback'' may arise from unitary Hamiltonian dynamics and from measurement backaction \cite{Nielsen2000,Wiseman2010}. 
While the analogy to the classical case sounds appealing, 
quantum mechanics imposes several fundamental differences. In particular, 
the probabilistic outcome of quantum measurements implies 
that the state dynamics will resemble a random walk where quantum jumps can cause large state changes. 
The present work explores the potential of such approaches in the context
of quantum state preparation for a system of $N$ qubits.
Our ideas may be tested in different platforms of current interest \cite{Bermudez2017,Arute2019,Minev2019,Satzinger2021,Mi2021,Monroe2021,Noel2022,WangGefen2022,Ferrergarcia2022,Wang2023a,Koh2023},
including superconducting qubit arrays, trapped ion setups, photonic circuits, and ultracold atom lattices.

Quantum state preparation in general represents a complex challenge 
\cite{Nielsen2000,Wiseman2010}.  One possible avenue is to employ deterministic schemes, without feedback from measurements over the course of the protocol.
For sufficiently small $N$, applying a sequence of single- and two-qubit
unitary gates (that is, a quantum circuit) is an option.  However, with increasing $N$, 
a deterministic gate sequence leading from a given initial state $|\Psi(0)\rangle$ 
to a predesignated and possibly highly entangled final target state $|\Psi_f\rangle$ is generally difficult to identify
and requires an exponential overhead of additional ancilla qubits \cite{Zhang2022a}.
If  $|\Psi_f\rangle$ represents a non-degenerate ground state of a gapped local Hamiltonian, one can resort to analog quantum simulations and/or quantum annealing or cooling methods \cite{Georgescu2014}, where the corresponding Hamiltonian is implemented in a controlled and tunable way, e.g., using lattices of 
trapped ions or ultracold atoms weakly coupled to a thermal environment.
For very low temperature, the quantum state then dynamically evolves towards the 
ground state $|\Psi_f\rangle$. 
However, such approaches may become impractical if the Hamiltonian  is gapless. Moreover,  a certain class of states cannot be expressed as ground states of a ``parent Hamiltonian,'' the latter being the sum of local Hamiltonians. Examples are exotic highly entangled $N$-qubit states such as the 
Green-Horne-Zeilinger (GHZ) state and the W state \cite{Nielsen2000,Dur2000}.
Written in the standard basis defined by Pauli-$Z$ operators ($\sigma^z|0\rangle=|0\rangle$ and $\sigma^z|1\rangle=-|1\rangle$ for the respective qubit),
\begin{eqnarray}\label{GHZn}
    |{\rm GHZ}\rangle &=& \frac{1}{\sqrt2}(|00\cdots 0\rangle+|11\cdots 1\rangle),\\ \nonumber
    |{\rm W}  \rangle &=& \frac{1}{\sqrt{N}} \left( |10\cdots 0\rangle + |01\cdots 0\rangle+ \cdots + |00\cdots 1\rangle\right),
\end{eqnarray}
represent two state classes with different types of high multipartite entanglement for $N>2$ qubits \cite{Dur2000}, where $|{\rm GHZ}\rangle$ (resp., $|{\rm W}\rangle$)  
is a superposition of two (resp., $N$) product states.

In this work, we instead study measurement-based strategies where the protocol is designed to autonomously find a desired target state of arbitrary form.   
It is widely recognized that quantum measurements offer great freedom in shaping and controlling the dynamics of quantum many-body systems, far beyond the possibilities offered by the unitary Hamiltonian dynamics of closed systems or 
the dissipative dynamics of driven open systems   \cite{Nielsen2000,Jacobs2006,Wiseman2010,Koch2016,Turkeshi2021,Harrington2022}.   
For instance, the interplay of projective measurements of stochastically selected qubits with the unitary dynamics due
to local two-qubit random gates causes nontrivial entanglement phase transitions in random quantum circuits, see Ref.~\cite{Fisher2023} for a recent review.
By allowing for projective measurements on detector qubits weakly coupled to the system qubits, weak measurement protocols can be designed for ``steering'' the system state $|\Psi(t)\rangle$ towards a desired target state $|\Psi_f\rangle$.  As customary in the recent literature\cite{Roy2020,Herasymenko2023,Ravindranath2023,Edd2023, Puente2024, Qian2024, Volya2024}, we here use the term ``steering'' as a proxy for ``guiding'' (or ``piloting'' \cite{Schroedinger1935}) the system state by a sequence of measurement processes.  We emphasize that this meaning is distinct from the notion of ``quantum steering'' in quantum information theory \cite{Wiseman2007,Uola2020,Xiang2022,Guhne2023}. In fact, we avoid the term ``quantum steering'' throughout.

One possibility is to employ \emph{passive} steering, where measurement outcomes are 
simply discarded \cite{Roy2020,Kumar2020,Kumar2022,Edd2023,Google2023}.  
 One may distinguish ``blind passive" from ``non-blind passive" steering,
where in the latter case the measurement readout is used to stop the protocol if it corresponds to a ``success" \cite{Volya2024}. However, the protocol is still passive since the steering operator is selected before the protocol started. 
Following Ref.~\cite{Herasymenko2023}, we refer to active steering only if a decision is made after 
each time step of the protocol. This decision depends on the history of previous measurement outcomes.
For many-body systems, non-blind passive protocols with local system-detector 
couplings are basically impossible since one cannot terminate the protocol based on the result of 
a single local measurement. While termination could be based on a long sequence of readouts, 
such a sequence may only be a probabilistic indicator of success. In addition, such approaches
typically suffer from the postselection problem (exponentially small chance of success) \cite{Fisher2023}.  

Importantly, there are major obstacles which can prevent the success of passive steering.  
First, while a state may in principle be passively steerable, the required steering operators can be so complicated that they are not available for the hardware at hand.    
Second, more fundamentally, it is impossible to passively steer certain state classes by \emph{any} set of \emph{local} steering operators. 
This is the case if the state cannot be represented as a ground state of a non-frustrated parent 
Hamiltonian \cite{Ticozzi2012}. The active steering protocols detailed in this work can overcome the no-go theorem of passive steering in 
Ref.~\cite{Ticozzi2012} for two reasons:  (i) In an active protocol, one follows the time evolution of the state and takes the decision (which steering operator is applied during the next time step) based on the history of previous measurement outcomes. (ii) Unlike with standard passive protocols, an important part of the state dynamics now comprises manipulations or measurements of the detector qubits. In our implementation, we introduce a
weak-measurement variant of entanglement swapping (for details, see below). 

It is worth emphasizing that previous applications of active decision making to the problem of steering, cf.\ Ref.~\cite{Herasymenko2023}, have only used sets of  ``passive'' measurement operators, i.e., operators which allow one to passively steer the system at long times. 
The goal was to accelerate steering by properly choosing --- based on the current system state --- one of the available operators at each step of the protocol. In the present work, a much more challenging question is addressed and solved: Is it possible to actively steer the system using a set of local measurement operators that would never yield the desired target state without active decisions?

Important examples for passively unsteerable (by local\footnote{We reiterate that the condition of locality here is crucial: the no-go theorem on steerability assumes steering operators acting only on a local subset of the system degrees of freedom (which does not scale up with system size). For example, in Ref.~\cite{Herasymenko2023}, passive steering of a three-qubit W state was addressed, which became possible because one of the steering operators used there involved all three qubits, thus representing ``global steering.'' In what follows, passively unsteerable states are understood as those that cannot be passively steered by applying local operators.} steering) $N$-qubit
states are
given by the GHZ and W states, see Eq.~\eqref{GHZn}. 
While tracing over a single qubit implies an only classically correlated reduced density matrix (RDM) for the remaining qubits for the GHZ state, 
the W state is more robust against entanglement loss.
Highly entangled states such as those in Eq.~\eqref{GHZn} can be used 
as a resource \cite{Chitambar2019}, e.g., for increasing the sensitivity of quantum detectors \cite{Giovannetti2011,Degen2017,Pirandola2018}.
We note that the GHZ state is a stabilizer state which can be prepared by projective measurements of a set of commuting products of Pauli operators (stabilizers)
augmented by Clifford operations or postselection if one measures ``wrong'' stabilizer eigenvalues \cite{Nielsen2000,Terhal2015}. 
Previous measurement-based state preparation experiments, see, e.g., Refs.~\cite{Riste2013,Roch2014,Chantasri2016,Blasiak2022}, have typically been
limited to stabilizer states.  Related ideas for preparing topologically ordered states in the large-$N$ limit have been described in recent works \cite{Briegel2001,Zeng2019,Roy2020,Lu2022,Verresen2023a,Verresen2023b,Bravyi2023,Lavasani2023,Zhu2023,Lee2023,Tanti2023,Smith2023,Wang2023a}. However, the underlying strategies apply only in special cases where, for instance, all measurement operators commute with all unitary gates. 

\begin{figure}[t]
\centering 
\footnotesize
\begin{tikzpicture}[scale=1.5]
            \node[] at (-0.5,1.5) (a) {\large (a)};
            \node[] at (2,1.8) (a) {\large $\mathbf{N=3}$};
                \node[state, draw=cyan!50, fill=cyan!50, minimum size=8mm] at (0,1.2) (a1) {$d_1$};
                \node[rectangle, draw=orange!60, fill=orange!60, minimum size=8mm] (s1) {$s_1$};
                \node[state, draw=cyan!50, fill=cyan!50, minimum size=8mm] at (1,1.2) (a2) {$d_2$};
                \node[rectangle, draw=orange!60, fill=orange!60, minimum size=8mm] at (1,0) (s2) {$s_2$};
                \node[state, draw=cyan!50, fill=cyan!50, minimum size=8mm] at (2,1.2) (a3) {$d_3$};
                \node[rectangle, draw=orange!60, fill=orange!60, minimum size=8mm] at (2,0) (s3) {$s_3$};

                \node[] at (0.5,1.85) {\includegraphics[scale=.35]{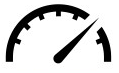}}; 
                
                \draw [-, black] (a1) edge[dashed, thick] node[left]{$H_1$} (s1);
                \draw [-, black] (a2) edge[dashed, thick] node[left]{$H_2$} (s2);
                \draw [-, black] (a3) edge[dashed, thick] node{} (s3);

                \draw [thick, fill=cyan!30] (0,1.55) to [out=150,in=30] (1,1.55) to [out=160,in=20] (0,1.55);
            \end{tikzpicture}
            
            \vspace{0.6cm}
            
            \begin{tikzpicture}[scale=1.8]
            \node[] at (0,1.5) (a) {\large (b)};
            \node[] at (2,-1.55) (a) {\large $\mathbf{N=6}$};
                \node[state, draw=cyan!50, fill=cyan!50, minimum size=3mm] at (0.5,1.2) (d11) {};
                \node[rectangle, draw=orange!60, fill=orange!60, minimum size=3mm] at (0.5,0.8) (s11) {};
                \draw [-, black] (d11) edge[thick] node[left]{} (s11);
                
                \node[state, draw=cyan!50, fill=cyan!50, minimum size=3mm] at (1,1.2) (d21) {};
                \node[rectangle, draw=orange!60, fill=orange!60, minimum size=3mm] at (1,0.8) (s21) {};
                \draw [-, black] (d21) edge[thick] node[left]{} (s21);
                
                \node[state, draw=cyan!50, fill=cyan!50, minimum size=3mm] at (1.5,1.2) (d31) {};
                \node[rectangle, draw=orange!60, fill=orange!60, minimum size=3mm] at (1.5,0.8) (s31) {};
                \draw [-, black] (d31) edge[thick] node[left]{} (s31);
                
                \node[state, draw=cyan!50, fill=cyan!50, minimum size=3mm] at (2,1.2) (d41) {};
                \node[rectangle, draw=orange!60, fill=orange!60, minimum size=3mm] at (2,0.8) (s41) {};
                \draw [-, black] (d41) edge[thick] node[left]{} (s41);

                \node[state, draw=cyan!50, fill=cyan!50, minimum size=3mm] at (2.5,1.2) (d51) {};
                \node[rectangle, draw=orange!60, fill=orange!60, minimum size=3mm] at (2.5,0.8) (s51) {};
                \draw [-, black] (d51) edge[thick] node[left]{} (s51);

                \node[state, draw=cyan!50, fill=cyan!50, minimum size=3mm] at (3,1.2) (d61) {};
                \node[rectangle, draw=orange!60, fill=orange!60, minimum size=3mm] at (3,0.8) (s61) {};
                \draw [-, black] (d61) edge[thick] node[left]{} (s61);
                
                \node[] at (0.75,1.5) {\includegraphics[scale=.25]{dial.png}};
                \draw [thick, fill=cyan!30] (0.5,1.325) to [out=150,in=30] (1,1.325) to [out=160,in=20] (0.5,1.325);
                \node[] at (1.75,1.5) {\includegraphics[scale=.25]{dial.png}};
                \draw [thick, fill=cyan!30] (1.5,1.325) to [out=150,in=30] (2,1.325) to [out=160,in=20] (1.5,1.325);
                \node[] at (2.75,1.5) {\includegraphics[scale=.25]{dial.png}}; 
                \draw [thick, fill=cyan!30] (2.5,1.325) to [out=150,in=30] (3,1.325) to [out=160,in=20] (2.5,1.325);
                
                \node[state, draw=cyan!50, fill=cyan!50, minimum size=3mm] at (0.5,0.2) (d12) {};
                \node[rectangle, draw=orange!60, fill=orange!60, minimum size=3mm] at (0.5,-0.2) (s12) {};
                \draw [-, black] (d12) edge[thick] node[left]{} (s12);
                
                \node[state, draw=cyan!50, fill=cyan!50, minimum size=3mm] at (1,0.2) (d22) {};
                \node[rectangle, draw=orange!60, fill=orange!60, minimum size=3mm] at (1,-0.2) (s22) {};
                \draw [-, black] (d22) edge[thick] node[left]{} (s22);
                
                \node[state, draw=cyan!50, fill=cyan!50, minimum size=3mm] at (1.5,0.2) (d32) {};
                \node[rectangle, draw=orange!60, fill=orange!60, minimum size=3mm] at (1.5,-0.2) (s32) {};
                \draw [-, black] (d32) edge[thick] node[left]{} (s32);
                
                \node[state, draw=cyan!50, fill=cyan!50, minimum size=3mm] at (2,0.2) (d42) {};
                \node[rectangle, draw=orange!60, fill=orange!60, minimum size=3mm] at (2,-0.2) (s42) {};
                \draw [-, black] (d42) edge[thick] node[left]{} (s42);

                \node[state, draw=cyan!50, fill=cyan!50, minimum size=3mm] at (2.5,0.2) (d52) {};
                \node[rectangle, draw=orange!60, fill=orange!60, minimum size=3mm] at (2.5,-0.2) (s52) {};
                \draw [-, black] (d52) edge[thick] node[left]{} (s52);

                \node[state, draw=cyan!50, fill=cyan!50, minimum size=3mm] at (3,0.2) (d62) {};
                \node[rectangle, draw=orange!60, fill=orange!60, minimum size=3mm] at (3,-0.2) (s62) {};
                \draw [-, black] (d62) edge[thick] node[left]{} (s62);
                
                \node[] at (0.25,0.5) {\includegraphics[scale=.25]{dial.png}};
                \draw [thick, fill=cyan!30] (0,0.325) to [out=150,in=30] (0.5,0.325) to [out=160,in=20] (0,0.325);
                \node[rectangle, draw=white, fill=white, minimum size=6mm] at (0.1,0.46) (dial1) {};
                \node[] at (1.25,0.5) {\includegraphics[scale=.25]{dial.png}};
                \draw [thick, fill=cyan!30] (1,0.325) to [out=150,in=30] (1.5,0.325) to [out=160,in=20] (1,0.325);
                \node[] at (2.25,0.5) {\includegraphics[scale=.25]{dial.png}};
                \draw [thick, fill=cyan!30] (2,0.325) to [out=150,in=30] (2.5,0.325) to [out=160,in=20] (2,0.325);
                \node[] at (3.25,0.5) {\includegraphics[scale=.25]{dial.png}};
                \draw [thick, fill=cyan!30] (3,0.325) to [out=150,in=30] (3.5,0.325) to [out=160,in=20] (3,0.325);
                \node[rectangle, draw=white, fill=white, minimum size=6mm] at (3.4,0.46) (dial2) {};

                \node[state, draw=cyan!50, fill=cyan!50, minimum size=3mm] at (0.5,-0.8) (d13) {};
                \node[rectangle, draw=orange!60, fill=orange!60, minimum size=3mm] at (0.5,-1.2) (s13) {};
                \draw [-, black] (d13) edge[thick] node[left]{} (s13);
                
                \node[state, draw=cyan!50, fill=cyan!50, minimum size=3mm] at (1,-0.8) (d23) {};
                \node[rectangle, draw=orange!60, fill=orange!60, minimum size=3mm] at (1,-1.2) (s23) {};
                \draw [-, black] (d23) edge[thick] node[left]{} (s23);
                
                \node[state, draw=cyan!50, fill=cyan!50, minimum size=3mm] at (1.5,-0.8) (d33) {};
                \node[rectangle, draw=orange!60, fill=orange!60, minimum size=3mm] at (1.5,-1.2) (s33) {};
                \draw [-, black] (d33) edge[thick] node[left]{} (s33);
                
                \node[state, draw=cyan!50, fill=cyan!50, minimum size=3mm] at (2,-0.8) (d43) {};
                \node[rectangle, draw=orange!60, fill=orange!60, minimum size=3mm] at (2,-1.2) (s43) {};
                \draw [-, black] (d43) edge[thick] node[left]{} (s43);

                \node[state, draw=cyan!50, fill=cyan!50, minimum size=3mm] at (2.5,-0.8) (d53) {};
                \node[rectangle, draw=orange!60, fill=orange!60, minimum size=3mm] at (2.5,-1.2) (s53) {};
                \draw [-, black] (d53) edge[thick] node[left]{} (s53);
                
                \node[state, draw=cyan!50, fill=cyan!50, minimum size=3mm] at (3,-0.8) (d63) {};
                \node[rectangle, draw=orange!60, fill=orange!60, minimum size=3mm] at (3,-1.2) (s63) {};
                \draw [-, black] (d63) edge[thick] node[left]{} (s63);
                
                \node[] at (0.75,-0.5) {\includegraphics[scale=.25]{dial.png}};
                \draw [thick, fill=cyan!30] (0.5,-0.675) to [out=150,in=30] (1,-0.675) to [out=160,in=20] (0.5,-0.675);
                \node[] at (1.75,-0.5) {\includegraphics[scale=.25]{dial.png}};
                \draw [thick, fill=cyan!30] (1.5,-0.675) to [out=150,in=30] (2,-0.675) to [out=160,in=20] (1.5,-0.675);
                \node[] at (2.75,-0.5) {\includegraphics[scale=.25]{dial.png}};
                \draw [thick, fill=cyan!30] (2.5,-0.675) to [out=150,in=30] (3,-0.675) to [out=160,in=20] (2.5,-0.675);
        
                \draw [->] (0.25,-1.35) -- node[left]{{\LARGE $t$}} (0.25,1.5);
                \draw [->] (0.3,-1.4) -- node{} (3.3,-1.4);
            \end{tikzpicture}
\caption{Cartoon of the active steering protocol. (a) $N$ system qubits (orange squares, here shown for $N=3$) are coupled  to the respective detector qubit (blue circles) by selected 
Pauli steering operators (here denoted by $H_1$ and $H_2$), see Eq.~\eqref{Pauligate} below. After a unitary time step, projective Bell measurements of neighboring detector qubit pairs  
(in this example for qubits $1$ and $2$, with qubit $3$ not being measured) implement weak local Bell measurements for
the state $|\Psi\rangle$ of the system qubits (entanglement swapping, see Fig.~\ref{fig2} for details).    
(b) Time evolution of the monitored quantum circuit with active feedback, shown for three time steps of a
chain of $N=6$ qubits with periodic boundary conditions.   A possible scheme for Bell measurements of neighboring detector qubit pairs  in subsequent cycles is 
indicated.  For details, see main text.}
\label{fig1}
\end{figure}

In this work, we show that passively unsteerable states, 
including non-stabilizer states such as the W state, 
can be prepared by active steering protocols using only a limited, and usually available, 
set of simple steering operators.  Even if $|\Psi_f\rangle$ is passively steerable, 
active steering will usually result in significantly faster protocols.
In general, active steering protocols involve decision making strategies 
where (a part of) the history of measurement outcomes is taken into account in the subsequent time evolution
\cite{Minev2019,Sivak2022,Liu2022a,Buchhold2023,Herasymenko2023,Friedman2023,Ravindranath2023,Wu2023,Hauser2023,Ravindranath2023a}.
 In our protocol, we consider $N$ system qubits each of which is allowed to
couple only to its own detector qubit partner, see Fig.~\ref{fig1}(a) for a schematic illustration. 
The respective system-detector coupling (``steering operator'') is chosen from a small 
set of Pauli operators, where the choice is determined by an active decision making strategy.  
We emphasize that direct couplings between system qubits are not needed and 
only a single steering operator (not superpositions thereof) is applied during a given time step of duration
$\delta t$.
One cycle then corresponds to a sequence of the following operations, see Fig.~\ref{fig1}(b): 
\begin{enumerate}
\item
One prepares the detector qubits in a simple product state. 
\item The chosen steering operators (see below for our active decision making policy) are switched on 
and unitary time evolution sets in. 
\item After the time $\delta t$ has elapsed, the 
steering operators are switched off again. One now performs 
projective Bell measurements on neighboring detector qubit pairs.
The measurement outcomes are used for selecting the 
steering operators in the next cycle.
\end{enumerate}
Since steering operators associated with different system qubits commute by construction,
and measurement operators for distinct (non-overlapping) qubit pairs commute as well, 
 one can simultaneously steer $[N/2]$ pairs in a given time step, cf.~Fig.~\ref{fig1}(b).  

\begin{figure}[t]
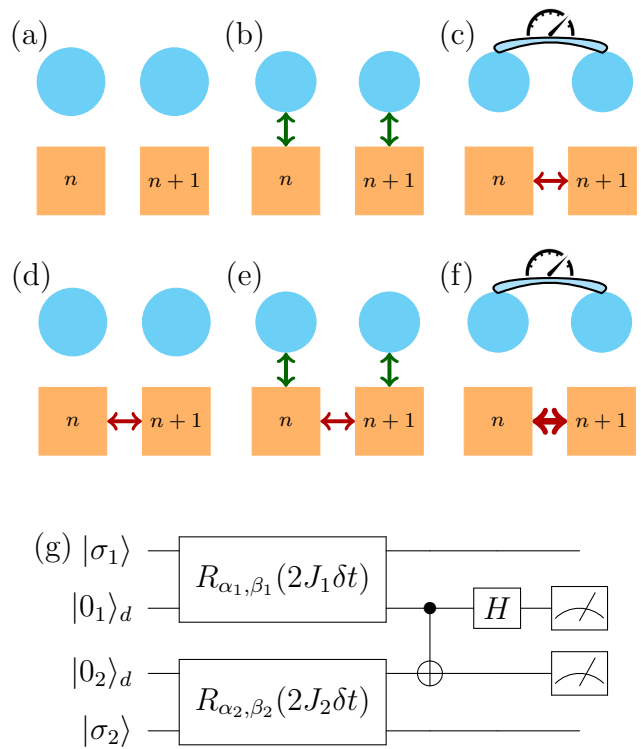

\centering 
\footnotesize
    \begin{tikzpicture}[scale=1.1]
    \node[] at (-0.5,1.75) (a) {\large (a)};
        \node[state, cyan!50, fill=cyan!50, minimum size=9mm] at (0,1.2) (a1) {};
        \node[rectangle, draw=orange!60, fill=orange!60, minimum size=9mm] (s1) {$n$};
        \node[state, cyan!50, fill=cyan!50, minimum size=9mm] at (1.25,1.2) (a2) {};
        \node[rectangle, draw=orange!60, fill=orange!60, minimum size=9mm] at (1.25,0) (s2) {$n+1$};

    \end{tikzpicture}
    \begin{tikzpicture}[scale=1.1]
    \node[] at (-0.5,1.75) (a) {\large (b)};
        \node[state, cyan!50, fill=cyan!50, minimum size=8mm] at (0,1.2) (a1) {};
        \node[rectangle, draw=orange!60, fill=orange!60, minimum size=9mm] (s1) {$n$};
        \node[state, cyan!50, fill=cyan!50, minimum size=8mm] at (1.25,1.2) (a2) {};
        \node[rectangle, draw=orange!60, fill=orange!60, minimum size=9mm] at (1.25,0) (s2) {$n+1$};

        \draw[<->, black!60!green, line width=1.5] (s1) -- node{} (a1);
        \draw[<->, black!60!green, line width=1.5] (s2) -- node{} (a2);
    \end{tikzpicture}
    \begin{tikzpicture}[scale=1.1]
    \node[] at (-0.5,1.75) (a) {\large (c)};
        \node[state, cyan!50, fill=cyan!50, minimum size=8mm] at (0,1.2) (a1) {};
        \node[rectangle, draw=orange!60, fill=orange!60, minimum size=9mm] (s1) {$n$};
        \node[state, cyan!50, fill=cyan!50, minimum size=8mm] at (1.25,1.2) (a2) {};
        \node[rectangle, draw=orange!60, fill=orange!60, minimum size=9mm] at (1.25,0) (s2) {$n+1$};

        \node[] at (0.625,1.9) {\includegraphics[scale=.35]{dial.png}}; 

        \draw[<->, black!30!red, line width=1.5] (s1) -- node{} (s2);
        \draw [thick, fill=cyan!30] (0,1.55) to [out=150,in=30] (1.25,1.55) to [out=160,in=20] (0,1.55);
    \end{tikzpicture}\\
     \vspace{0.3cm}
    
    \begin{tikzpicture}[scale=1.1]
    \node[] at (-0.5,1.75) (a) {\large (d)};
        \node[state, cyan!50, fill=cyan!50, minimum size=9mm] at (0,1.2) (a1) {};
        \node[rectangle, draw=orange!60, fill=orange!60, minimum size=9mm] (s1) {$n$};
        \node[state, cyan!50, fill=cyan!50, minimum size=9mm] at (1.25,1.2) (a2) {};
        \node[rectangle, draw=orange!60, fill=orange!60, minimum size=9mm] at (1.25,0) (s2) {$n+1$};

        \draw[<->, black!30!red, line width=1.5] (s1) -- node{} (s2);
    \end{tikzpicture}
    \begin{tikzpicture}[scale=1.1]
    \node[] at (-0.5,1.75) (a) {\large (e)};
        \node[state, cyan!50, fill=cyan!50, minimum size=8mm] at (0,1.2) (a1) {};
        \node[rectangle, draw=orange!60, fill=orange!60, minimum size=9mm] (s1) {$n$};
        \node[state, cyan!50, fill=cyan!50, minimum size=8mm] at (1.25,1.2) (a2) {};
        \node[rectangle, draw=orange!60, fill=orange!60, minimum size=9mm] at (1.25,0) (s2) {$n+1$};

        \draw[<->, black!30!red, line width=1.5] (s1) -- node{} (s2);
        \draw[<->, black!60!green, line width=1.5] (s1) -- node{} (a1);
        \draw[<->, black!60!green, line width=1.5] (s2) -- node{} (a2);
    \end{tikzpicture}
    \begin{tikzpicture}[scale=1.1]
    \node[] at (-0.5,1.75) (a) {\large (f)};
        \node[state, cyan!50, fill=cyan!50, minimum size=8mm] at (0,1.2) (a1) {};
        \node[rectangle, draw=orange!60, fill=orange!60, minimum size=9mm] (s1) {$n$};
        \node[state, cyan!50, fill=cyan!50, minimum size=8mm] at (1.25,1.2) (a2) {};
        \node[rectangle, draw=orange!60, fill=orange!60, minimum size=9mm] at (1.25,0) (s2) {$n+1$};

        \node[] at (0.625,1.9) {\includegraphics[scale=.35]{dial.png}}; 

        \draw[<->, black!30!red, line width=2.5] (s1) -- node{} (s2);
        \draw [thick, fill=cyan!30] (0,1.55) to [out=150,in=30] (1.25,1.55) to [out=160,in=20] (0,1.55);
    \end{tikzpicture}\\
    \vspace{0.3cm}
\[
 \text{\large (g)}\hspace{1cm}\large
 \Qcircuit @C=1em @R=0.7em {
        \lstick{|\sigma_1\rangle} & \multigate{1}{R_{\alpha_1,\beta_1}(2J_1\delta t)} & \qw & \qw & \qw\\
        \lstick{|0_{1}\rangle_{d} }& \ghost{R_{\alpha_1,\beta_1}(2J_1\delta t)} & \ctrl{1} & \gate{H} & \meter\\
        \lstick{|0_{2}\rangle_d} & \multigate{1}{R_{\alpha_2,\beta_2}(2J_2\delta t)} & \targ & \qw & \meter\\
        \lstick{|\sigma_2\rangle} & \ghost{R_{\alpha_2,\beta_2}(2J_2\delta t)} & \qw & \qw & \qw
}
\]
\caption{ 
Schematic illustration of entanglement generation during one time step of the protocol. We use
a weak-measurement variant of entanglement swapping \cite{Boschi1998,Pan1998,Jennewein2001,Gisin2005,Riebe2008,Kaltenbaek2009,Horodecki2009,Huang2023}, where one steers the neighboring system qubits ($n,n+1$) indicated by squares.  
Circles depict the corresponding detector qubits as in Fig.~\ref{fig1}.  
(a) Configuration at initial time $t=0$, where all qubits are decoupled from each other. (b) At time $\delta t$ (after one time step), each system-detector qubit pair is
entangled (green vertical arrows) due to unitary time evolution under the chosen steering operators.
(c) After a projective Bell measurement of the detector pair, system and detector qubits disentangle and, depending on the measurement outcome, 
entanglement between the system qubits (red horizontal arrow) builds up.
Panels (d)-(f) show the corresponding steps during the subsequent time step of the protocol, where entanglement generated in the previous time step is already
present.  Typically, entanglement increases after the Bell measurement (thicker red horizontal arrow).  
(g) Quantum circuit representation of  entanglement swapping. The Pauli rotations ($R$) depend on the chosen steering operators (see Sec.~\ref{sec2a} for details) and act on system and detector qubits, with a subsequent measurement of the detector qubits in the Bell basis. System (detector) qubit states are here denoted by $|\sigma_{1,2}\rangle$ 
 ($|\sigma_{1,2}\rangle_d$).  The Bell basis measurement is achieved by CNOT and Hadamard gates.   } 
\label{fig2}
\end{figure}

The above Bell measurements realize weak measurements on the system qubits and provide an efficient way to systematically generate 
entanglement in $|\Psi\rangle$ by means of a weak-measurement version of  entanglement swapping  \cite{Boschi1998,Pan1998,Jennewein2001,Gisin2005,Riebe2008,Kaltenbaek2009,Horodecki2009,Huang2023}, where 
entanglement is incrementally teleported from the detector qubits to the system qubits.
(We simply refer to ``entanglement swapping'' in what follows.)
We illustrate the basic mechanism in Fig.~\ref{fig2}, 
where $|\sigma_{n,n+1}\rangle$ ($|\sigma_{n,n+1}\rangle_d$) with $\sigma=0,1$ refer to basis states of the 
respective system (detector) qubits $n$ and $n+1$.  In panels (a)-(c), we consider the first cycle of the protocol, starting at the initial time $t=0$: 
(a) One starts from the simple product state $(|0_n\rangle\otimes |0_n\rangle_d) \otimes (|0_{n+1}\rangle\otimes |0_{n+1}\rangle_d)$. 
 (b) Switching on the selected steering operators during a time step of duration $\delta t$, each system and detector qubit pair are entangled 
  by the unitary time evolution. (c) The subsequent projective Bell measurement of the detector qubits then disentangles the system and detector qubits but 
  at the same time generates entanglement among the system qubits by means of entanglement swapping.  We provide a detailed discussion of this step in 
  Appendix~\ref{app0}. Next, in panels (d)-(f) we illustrate what happens during the next cycles of the protocol: 
  (d) As a result of the last step, some entanglement may now be present in the system state $|\Psi(\delta t)\rangle$, before one applies the 
  steering operators for the next step. (e) After the unitary time step, each system-detector qubit pair becomes entangled. (f) One performs again a Bell measurement, 
  which typically increases the entanglement in the resulting state $|\Psi(2\delta t)\rangle$. 
  Finally, we provide a quantum circuit representation of this entanglement swapping protocol
  \cite{Nielsen2000} 
  in Fig.~\ref{fig2}(g).   A detailed discussion of our entanglement swapping protocol can be found in Sec.~\ref{sec2a} and in Appendix~\ref{app0}.

We emphasize that the re-initialization of detector qubits after the measurements is possible in our scheme, see also Refs.~\cite{Koh2023,Volya2024}. Without such a property, the number of detector qubits would  proliferate for large circuit depth.  Moreover,
since the implementation of such measurements only requires two-qubit interactions, we expect that  entanglement swapping is easier to realize experimentally than, for instance, the generation of entanglement through measurements done on  single detector qubits coupled to two or more system qubits.  In fact, one can show  
that in the latter setup, with a fixed set of Pauli steering operators, it is not possible to reach arbitrary target states $|\Psi_f\rangle$, not
even for $N=2$.  We provide a discussion of this point in Appendix~\ref{app1}. 
As shown below, entanglement swapping does not suffer from this problem. 
Measuring a pair of qubits in their Bell basis also is a standard procedure in many platforms of current interest.
More generally, we find that an arbitrary state obtainable by consecutively entangling pairs of qubits, say, by applying two-qubit gates, can also be engineered by means of active steering based on entanglement swapping. In principle, this includes highly entangled states with volume-law growth of the entanglement entropy. 

In order to make progress, we consider an idealized situation free from external noise or other
imperfections \cite{Edd2023}.  In particular, we assume that the parameters characterizing allowed
steering operators are known from prior calibration runs.  If the initial pure state $|\Psi(0)\rangle$ (e.g., an easily accessible product state) and the measurement outcomes are known,   
one can then track the state trajectory $|\Psi(t)\rangle$ at later times $t=m\delta t$ (integer $m$) by a classical 
calculation. (For a given measurement record,  $|\Psi(t)\rangle$ always remains  
pure. By averaging over measurement outcomes, one may obtain a mixed state.)

For moderate values of $N$, those assumptions may be justified to reasonable accuracy on time scales below the respective qubit dephasing time.
This case is also of practical interest if parts of the system can be steered locally and the whole system is then composed of small-$N$ blocks, see Ref.~\cite{Smith2023} for related work.
A different perspective  is to run our protocol on a fault-tolerant circuit with quantum error correction \cite{Nielsen2000,Terhal2015,Andersen2020,Stricker2020,Egan2021,Ryan2021,Krinner2022,Abobeih2022,Zhao2022a,Google2022}, where all system and detector qubits are encoded logical qubits.  Errors related to external noise can thereby be 
detected and actively corrected.  We note that recent experimental works
managed to beat the break-even point for fault tolerance of a logical qubit \cite{Sivak2023,Ni2023}.
In Sec.~\ref{sec4c},  potential extensions of our protocol 
to noisy quantum circuits will be discussed, see also Ref.~\cite{Edd2023}.  
While we show numerical simulation results based on the
present formulation of our active steering protocol only for $N\le 6$ system qubits in
Sec.~\ref{sec3}, we expect that large values of $N$ can be reached using
closely related protocols, see Sec.~\ref{sec4B}. 
Our work represents a proof of concept that active steering protocols can reach 
passively unsteerable target states. 

A central element of our approach is active feedback: based on the 
measurement outcomes recorded after a given time step, the steering operators used in  
the next time step are determined by optimizing a non-negative cost function $C(t)$.  Convergence, i.e., 
$|\Psi\rangle=e^{i\gamma}|\Psi_f\rangle$ with arbitrary global phase $\gamma$, 
is reached only for $C=0$, such that one has to minimize $C(t)$. In practice,
for each steering parameter configuration $K$,
we compute the expected cost function change in the next time step,
\begin{equation}\label{dck}
    \overline{dC(K)} \equiv \overline{C(t+\delta t; K)} - C(t),
\end{equation}
where the overbar indicates an average over measurement outcomes.
The steering operator $K$ is then chosen to maximize the cost function gain $|\overline{dC(K)}|$ with $\overline{dC(K)}\le 0$.  
For a proper cost function, and assuming a sufficiently large (``universal'') steering operator set, 
$\overline{dC(K)}<0$ is required for at least one $K$ unless one has already reached convergence. 
An obvious  cost function candidate is related to the fidelity $F$ \cite{Nielsen2000}, 
\begin{equation}\label{cost}
    C_N= 1-F^2,\quad F= |\langle \Psi_f|\Psi\rangle|.
    \end{equation}
However, Eq.~\eqref{cost} is \emph{not} sufficient for our active steering protocol. 
For instance, consider steering from $|\Psi\rangle=|00\cdots 0\rangle$ to $|\Psi_f\rangle
=|11\cdots 1\rangle$.  
Flipping  single (or a few) qubits, $|0\rangle\to |1\rangle$, evidently brings one closer to the
target state yet the resulting state remains orthogonal, $F=0$, 
unless one flips all $N$ qubits simultaneously.  In essence, we encounter a situation reminiscent of 
Anderson's orthogonality catastrophe \cite{Anderson1967}  with similarities to
the barren plateau problem in machine learning \cite{Cerezo2021}. 
In any case, when using Eq.~\eqref{cost} as cost function,  active steering will 
generally fail.  This failure already affects the minimal case $N=2$, see Sec.~\ref{sec2c} for details, 
and becomes more severe with increasing $N$.  
We show below that by incorporating ``local'' fidelity cost function terms, this problem can fortunately be resolved. (``Locality'' is here meant 
in the sense of operator locality.) For all possible RDMs for $r<N$ qubits built from $|\Psi(t)\rangle$, 
those terms minimize a suitable operator distance measure between the RDM and the respective RDM constructed from $|\Psi_f\rangle$. We address convergence properties of the resulting active 
steering protocols in Sec.~\ref{sec2d}. 

The structure of the remainder of this paper is as follows. In Sec.~\ref{sec2}, we discuss the detailed quantum state dynamics
of the actively monitored circuit described above.  In Sec.~\ref{sec3}, we present numerical simulation results to illustrate 
the performance of the active steering protocol. 
We then discuss implementation aspects and summarize possible perspectives 
for future work in Sec.~\ref{sec4}.  
Technical details can be found in the Appendix. Throughout, we use units with $\hbar=1$.

\section{Active steering protocol}\label{sec2}

We now turn to a quantitative description of the active steering protocol.  
The measurement-conditioned dynamics of the quantum circuit in Fig.~\ref{fig1}
is analyzed in detail in Sec.~\ref{sec2a}. For concreteness, we study 
a chain of $N$ system qubits with periodic boundary conditions, 
where each system qubit can be coupled to an detector qubit.
However, our protocol can easily be adapted to other circuit geometries.
 In Sec.~\ref{sec2b}, we summarize the Bloch tensor
state representation which considerably simplifies the subsequent steps.
We proceed with an analysis of the active decision making strategy in Sec.~\ref{sec2c}, which crucially depends on the construction of a cost function employed in the optimization process. Utilization of a global fidelity cost function introduces ``weak values'' \cite{Aharonov1988} into the ensuing stochastic evolution equation.  Using these weak
values, we show that  successful active steering is only possible if local fidelity terms are added to the cost function. Finally, we discuss convergence aspects in Sec.~\ref{sec2d}.
Additional details can be found in Appendix~\ref{app2}.

\subsection{Stochastic Schr\"odinger equation}\label{sec2a}

In what follows, system qubits are described by the operators $\sigma_n^{\mu}$, 
where $n=1,\ldots,N$ labels the qubits, $\mu=0$ corresponds to the identity, and 
for $\mu=\alpha\in \{1,2,3\}=\{x,y,z\}$, we have standard Pauli matrices.  
Similarly, detector qubits correspond to the operators $\tau_n^\mu$. 
For clarity, we assume below that all states of the uncoupled system qubits and, separately, 
all states of the detector qubits are degenerate, assigning to them zero energy. 
In other words, we assume no intrinsic Hamiltonian of the system and detector qubits.
During a unitary time evolution step, a steering operator 
$H_{n,K_n}$ can couple the system qubit $\sigma_n$ with the corresponding detector qubit $\tau_n$,  
where $K_n$ is chosen from a set of allowed steering parameters.
We here consider Pauli steering operators,
\begin{equation}\label{Pauligate}
    H_{n,K_n}= s_n J_n \sigma_n^{\alpha_n} \tau_n^{\beta_n},
\end{equation}
with the sign $s_n\in\{1,-1\}$ and the Pauli operator 
indices $\alpha_n\in\{x,y,z\}$ and $\beta_n\in \{x,y,z\}$. (Identity operators acting on other qubits are often kept implicit below.)
Clearly, the steering operators in Eq.~\eqref{Pauligate} commute for $n\ne n'$.
We stress that for each pair $(\sigma_n,\tau_n)$, only a single Pauli operator couples the 
respective system and detector qubit 
during a given time step --- no superpositions of different Pauli operators are needed in our protocol.
The coupling constants $J_n>0$ are assumed to be fixed and known, such that the available steering operators 
are parametrized by $K_n=(s_n,\alpha_n,\beta_n)$. 
Without loss of generality, the sign $s_n$ can be restricted to $s_n=+1$ for $\beta_n\ne z$,
implying a set of 12 possible steering parameters $\{K_n\}$.  Further reductions of the steering operator set are possible in special 
cases.  For instance, we empirically find that in order to prepare the GHZ state, one can exclude steering operators with $\beta_n=y$. Such operators
are, however, needed for the W state.  Determining the minimal set of steering operators required for achieving convergence to general target states
remains an interesting open issue.

Let us now consider a specific nearest-neighbor qubit pair ($n,n+1$) as illustrated in Fig.~\ref{fig2}. Before each unitary time 
evolution step, the corresponding detector qubits are initialized in a simple product state, say, $|00\rangle_d$. (The index $d$ distinguishes the detector Hilbert space from system space.)
Subsequently, the steering operator $H_K = H_{n,K_n}+H_{n+1,K_{n+1}}$ is switched on for suitably chosen steering parameters $K=(K_n,K_{n+1})$, 
and unitary time evolution of the full system sets in.  After the time step $\delta t$ with $J_{n/n+1}\delta t\ll 1$,
the steering operator $H_K$ is switched off again, and one projectively measures the detector pair in its Bell basis \cite{Nielsen2000},
\begin{eqnarray}\label{bell}
|\Phi_{\xi=0,\eta=\pm}\rangle_d &=& \left(|00\rangle_d \pm |11\rangle_d\right) /\sqrt2,\\
\quad |\Phi_{\xi=1,\eta=\pm}\rangle_d &=& \left(|01\rangle_d \pm |10\rangle_d\right)/\sqrt2,\nonumber
\end{eqnarray}
where symmetric ($\eta=+1$) and antisymmetric ($\eta=-1$) Bell states have either 
even $(\xi=0)$ or odd ($\xi=1$) parity.  Here ``symmetry'' refers to the behavior of the state under exchange of the two qubits,
while even (odd) ``parity'' means that it is built from the basis states $\{ |00\rangle_d, |11\rangle_d\}$  ($\{|01\rangle_d, |10\rangle_d$\}).
Since the initial state $|00\rangle_d$ obviously has even parity, we refer to measurement outcomes with $\xi=1$ as ``quantum jumps''. 
By means of entanglement swapping \cite{Boschi1998,Pan1998,Jennewein2001,Gisin2005,Riebe2008,Kaltenbaek2009,Horodecki2009,Huang2023}, see Fig.~\ref{fig2} and Appendix~\ref{app0},
one can then generate entanglement in the system state $|\Psi\rangle$, where the binary stochastic variables $\xi$ and $\eta$  encode the measurement outcomes. Their probabilistic nature is a fundamental consequence of quantum mechanics.
We emphasize that the above Bell measurements are local in the sense that only nearest-neighbor detector pairs are probed.  
In practice, they can be implemented by simultaneously measuring two commuting Pauli operators for detector qubit pairs \cite{Nielsen2000}, 
\begin{equation}\label{POp}
    \hat{\cal O}^x=\tau_n^x \tau_{n+1}^x, \quad \hat {\cal O}^z=\tau_n^z\tau_{n+1}^z.
\end{equation}
Indeed, one easily checks that 
\begin{equation}
    {\cal O}^x|\Phi_{\xi,\eta}\rangle_d= \eta |\Phi_{\xi,\eta}\rangle_d. \quad {\cal O}^z|\Phi_{\xi,\eta}\rangle_d= (1-2\xi)|\Phi_{\xi,\eta}\rangle_d.
\end{equation} 
A projective measurement of the Bell state $|\Phi_{\xi,\eta}\rangle_d$ thus implies that one measures the eigenvalues ${\cal O}^x=\eta(=\pm 1)$ and
${\cal O}^z=+1$ (for $\xi=0$) or ${\cal O}^z=-1$ (for $\xi=1$) for the observables in Eq.~\eqref{POp}.
Finally, we note that for the quantum circuit in Fig.~\ref{fig2}(g),
 in the computational basis of the two detector qubits, the measurement has the possible outcomes $|\frac{1-\eta}{2}, \xi\rangle_d$.

For measurement outcome ($\xi,\eta$) after a completed time step, the state of the full system is
given by $|\Psi(t+\delta t)\rangle \otimes |\Phi_{\xi,\eta}\rangle_d$, where the system state 
$|\Psi(t+\delta t)\rangle$ depends on $(\xi,\eta)$ because of  entanglement swapping, see Fig.~\ref{fig2} and Appendix~\ref{app0}.
In our protocol, since direct couplings between system qubits are absent, 
entanglement is solely generated (or removed) through the above  Bell measurements. Upon 
re-initializing the detector pair, the full state is prepared for the next
time step as $|\Psi(t+\delta t)\rangle\otimes |00\rangle_d$.  
Since all measurement operators $\hat {\cal O}^{x,z}$ for non-overlapping pairs commute, and also the corresponding steering operators $H_K$ commute, one can simultaneously steer and measure $[N/2]$ distinct pairs in one time step. For instance, for $N=4$ qubits, we could steer the qubit pairs $(1,2)$ and $(3,4)$ in the first cycle, and then the pairs $(2,3)$ and $(4,1)$ in the second cycle.  We do not allow for next-nearest-neighbor couplings within our notion of locality, i.e., only adjacent detector qubit pairs are subject to Bell measurements.  
Hence direct pair-steering of  $(1,3)$ and $(2,4)$ is not taken into account.
In general, for subsequent time steps, one can choose either a random pattern or an alternating pattern of measurement pairs;
the latter case is shown in Fig.~\ref{fig1}(b).  We find that a randomized pair assignment gives 
slightly superior convergence properties, but both schemes work well in practice.

Given the above remarks, it suffices to analyze the state change in one time step
when steering and measuring only a single pair $(n,n+1)$.  
For the measurement outcome $(\xi,\eta)$, the state $|\Psi(t+\delta t)\rangle$ after the measurement
is given by \cite{Nielsen2000,Jacobs2006,Wiseman2010}
\begin{equation}\label{Krausdef}
    |\Psi(t+\delta t)\rangle = \frac{1}{\sqrt{P_{\xi,\eta}}} A_{\xi,\eta} |\Psi(t)\rangle,   
\end{equation}
with the Kraus operators $A_{\xi,\eta}={}_d\langle \Phi_{\xi,\eta}| e^{-i\delta t H_K}|00\rangle_d$ and the outcome probabilities  
\begin{equation}\label{Pdef}
    P_{\xi,\eta}=\langle \Psi(t)| A_{\xi,\eta}^\dagger A_{\xi,\eta}^{}|\Psi(t)\rangle.
\end{equation}
In the weak-measurement limit \cite{Wiseman2010}, 
the couplings $J_n$ and the time step $\delta t$ are adjusted such that the rates
\begin{equation}\label{gammadef}
    \Gamma_n=J_n^2\delta t
\end{equation} 
are effectively independent of $\delta t$ and satisfy $J_n\delta t=\sqrt{\Gamma_n\delta t}\ll 1$. 
We assume this limit throughout.

Expanding the Kraus operators to lowest nontrivial order in $\delta t$, we then obtain
\begin{equation}\label{Adef}
    A_{\xi,\eta} = \sqrt{\frac{\delta t}{2}} \xi c_\eta + \frac{1}{\sqrt{2}} (1-\xi) \left[ 1 - i\delta t\left(
    H_{\eta} -\frac{i}{2} c^\dagger_\eta c_\eta^{}  \right)\right],
\end{equation}
where the effective Hamiltonian $H_{\eta}$ and the jump operators $c_{\eta}$ act on the system Hilbert space.
These quantities apply to a given time step and for the chosen qubit pair ($n,n+1$).  We note that
$H_{\eta}$ only appears in Eq.~\eqref{Adef} if one measures the same (even, $\xi=0$) parity of 
$|\Phi_{\xi,\eta}\rangle_d$ as for the initial detector state $|00\rangle_d$. 
With the Kronecker symbol $\delta_{\beta,\alpha}$ ($\alpha=x,y,z$), we obtain
\begin{eqnarray}\label{cc}
   && H_{\eta=\pm} = \sum_{m=n,n+1} s_m J_m \delta_{\beta_m,z} \sigma_m^{\alpha_m} +\\ 
    \nonumber
   && \, +\, \eta\sqrt{\Gamma_n\Gamma_{n+1}} (\delta_{\beta_n,x}\delta_{\beta_{n+1},y}+
   \delta_{\beta_n,y}\delta_{\beta_{n+1},x}) \sigma_n^{\alpha_n}\sigma_{n+1}^{\alpha_{n+1}},
\end{eqnarray}
where the term $\propto \eta$ arises from a Lamb shift contribution.
In addition, for $\beta_m\ne z$,  quantum jumps --- transitions from $|00\rangle_d$ into the odd-parity sector signaled by $\xi=1$ --- are possible, with the respective jump rates given by 
Eq.~\eqref{gammadef}.
With the compact notation 
\begin{equation}
    \delta^{(c)}_{\beta,\perp}=\delta_{\beta,x}+i\delta_{\beta,y}, \quad \delta_{\beta,\perp}=\delta_{\beta,x}+\delta_{\beta,y},
\end{equation}
the jump operators take the form ($\eta=\pm$)
\begin{equation}\label{jumpops}
   c_\eta = -i\left( \eta  \sqrt{\Gamma_n} \delta^{(c)}_{\beta_n,\perp} \sigma_n^{\alpha_n}  +  \sqrt{\Gamma_{n+1}}\, \delta^{(c)}_{\beta_{n+1},\perp} \sigma_{n+1}^{\alpha_{n+1}}\right).
\end{equation}
From Eq.~\eqref{Adef}, the outcome probabilities \eqref{Pdef} can be written as
\begin{equation}\label{Pdef2}
    P_{\xi,\eta} =  \frac12 \left( \delta_{\xi,0} + (\delta_{\xi,1}-\delta_{\xi,0}) \delta t\,  \langle c^\dagger_\eta c_\eta^{} \rangle \right) 
\end{equation}
with $\langle c^\dagger_\eta c_\eta^{} \rangle = \langle\Psi(t)| c^\dagger_\eta c_\eta^{} |\Psi(t)\rangle.$
From Eq.~\eqref{jumpops}, we obtain
\begin{eqnarray}\nonumber
    \langle c^\dagger_\eta c^{}_\eta \rangle &=&  2\eta\sqrt{\Gamma_n\Gamma_{n+1}} \left( \delta_{\beta_n,x}\delta_{\beta_{n+1},x}+
    \delta_{\beta_n,y}\delta_{\beta_{n+1},y}\right) \times\\
    &\times& Q_{n,n+1}^{\alpha_n,\alpha_{n+1}}  + \sum_{m=n,n+1}\Gamma_m\delta_{\beta_m,\perp} \label{avcdc}
\end{eqnarray}
with the   correlation function
\begin{equation}\label{spincor}
Q_{n,n'}^{\alpha,\alpha'}  = \langle\Psi(t)| \sigma_n^{\alpha} \sigma_{n'}^{\alpha'} |\Psi(t)\rangle . 
\end{equation}
The average of the binary quantum jump variable $\xi\in \{0,1\}$ then has a state-independent form,
\begin{equation}\label{bareta}
\overline{\xi} = \sum_{\xi,\eta} \xi P_{\xi,\eta} = \sum_{m=n,n+1} \Gamma_m \delta t\,\delta_{\beta_m,\perp} . 
\end{equation}
Since $\overline{\xi}\propto \delta t$, terms $\propto \xi\delta t$ are beyond 
the accuracy of the lowest-order expansion in $\delta t$ and will be neglected throughout.  However, the identity $\xi^2=\xi$ implies that terms involving products of $\xi$ are of order $\delta t$ and must be retained \cite{Jacobs2006,Wiseman2010}.

From Eqs.~\eqref{Krausdef} and \eqref{Adef}, we find that the state change after one time step, 
$|d\Psi\rangle= |\Psi(t+\delta t)\rangle- |\Psi(t)\rangle$, follows from
a jump-type nonlinear stochastic Schr\"odinger equation
\cite{Jacobs2006,Breuer2002,Wiseman2010,Zhang2017},
\begin{eqnarray} \nonumber
|d\Psi\rangle &=&\Biggl [ -i\delta t H_\eta+\xi \Biggl( \frac{c_{\eta}^{}}{\sqrt{\langle c_{\eta}^\dagger c_{\eta}^{} \rangle}}-1\Biggr)  \\  \label{SSE} && \qquad 
- \frac{\delta t}{2}  \left(c^\dagger_{\eta} c^{}_{\eta} - \langle c_{\eta}^\dagger c_{\eta}^{} \rangle\right) 
\Biggr] |\Psi(t)\rangle.
\end{eqnarray}
Unless noted otherwise, we assume an easily accessible pure product initial state, say, 
$|\Psi(0)\rangle=|00\cdots 0\rangle$.  For  given measurement record, the state 
$|\Psi(t)\rangle$ then always remains pure and follows from Eq.~\eqref{SSE}.
Depending on the choice of the steering operators, the quantum state dynamics in 
Eq.~\eqref{SSE} includes single- and two-qubit unitary Hamiltonian terms 
(from $H_\eta$), single-qubit Pauli gates (due to quantum jumps for $\xi=1$), and  
two-qubit unitary terms (from the last term which represents state diffusion \cite{Jacobs2006}).  In principle, given the set of steering operators in Eq.~\eqref{Pauligate} which in turn determine
the jump operators $c_\eta$ and the Hamiltonian $H_\eta$,
all operations needed for accessing a target state of arbitrary form 
are therefore contained in Eq.~\eqref{SSE} \cite{DiVincenzo1995,Lloyd1995,Nielsen2000,Marvian2022}.
We discuss convergence properties of our protocol in Sec.~\ref{sec2d}.

One can equivalently describe the state dynamics by
using the density matrix, $\rho(t)=|\Psi(t)\rangle\langle\Psi(t)|$. Using Eq.~\eqref{SSE}, 
the change in one time step,  $d\rho=\rho(t+\delta t)-\rho(t),$ is determined by the stochastic master equation \cite{Wiseman2010}
\begin{equation}\label{drho2}
 d\rho = -i\delta t [ H_\eta,\rho ]+ \xi\Biggl( \frac{c_{\eta} \rho c_{\eta}^\dagger}{\langle c_{\eta}^\dagger c_{\eta}^{}\rangle}  -
 \rho \Biggr) -\frac{\delta t}{2} \left \{ c^\dagger_{\eta} c^{}_{\eta}-\langle c^\dagger_{\eta} c^{}_{\eta}\rangle,\rho\right\}, 
 \end{equation}
with the anticommutator $\{ \cdot ,\cdot \}$.  State normalization holds because of ${\rm Tr}(d\rho)=0$.
Apart from a unitary evolution term due to $H_\eta$, Eq.~\eqref{drho2} again contains a jump term $\propto \xi$ which can result in large state changes, and a stochastic contribution of diffusion type (the anticommutator). 

Averaging Eq.~\eqref{drho2} over measurement outcomes using the probabilities \eqref{Pdef2} 
will usually result in a mixed state. The averaged state change, $\overline{d\rho}=\sum_{\xi,\eta} 
d\rho_{\xi,\eta} P_{\xi,\eta}$ with $d\rho_{\xi,\eta}$ in Eq.~\eqref{drho2}, 
is governed by a Lindblad master equation \cite{Breuer2002,Wiseman2010},
\begin{eqnarray}\nonumber
    \overline{d\rho} &=& \delta t\sum_{m=n,n+1} \Bigl( - i s_m J_m\delta_{\beta_m,z}  [\sigma_m^{\alpha_m},\rho] + \\ \label{ddrho2}
   & &\qquad \qquad +\, \Gamma_m \delta_{\beta_m,\perp} {\cal L}[\sigma_m^{\alpha_m}] \rho \Bigr),
\end{eqnarray}
with ${\cal L}[\sigma_m^{\alpha_m}]\rho= \sigma_m^{\alpha_m}\rho\sigma_m^{\alpha_m}-\rho$.
Equation \eqref{ddrho2} corresponds to the time evolution under a blind steering protocol, 
where measurement outcomes are simply discarded.
Since  only uncorrelated contributions from qubits $n$ and $n+1$ appear in Eq.~\eqref{ddrho2}, the
averaged state dynamics by itself is not capable of detecting entanglement structures 
built up in measurement-conditioned state trajectories. 

\subsection{Bloch tensor representation}\label{sec2b}

Given the Pauli form of the steering operators \eqref{Pauligate}, it is
convenient to employ the rank-$N$ Bloch tensor $R_{{\cal S}}$ for representing states.   
(In quantum information theory, $R_{{\cal S}}$ is usually referred to as $N$-qubit Bloch vector.) 
To that end, we first define Pauli string operators of length $N$,
\begin{equation}\label{string}
    {\cal S} = \sigma_1^{\mu_1} \sigma_2^{\mu_2} \cdots \sigma_N^{\mu_N},\quad \mu_j\in\{0,1,2,3\}.
\end{equation}
(We synonymously use the notations $\mu_j= x,y,z$ and $\mu_j=1,2,3$.)
An arbitrary $N$-qubit state $\rho$ can be written in the form
\begin{equation}\label{Bloch}
  \rho = \frac{1}{2^N} \sum_{\cal S} R_{\cal S} {\cal S}, 
  \quad R_{\cal S}={\rm Tr} (\rho {\cal S}),
\end{equation}
where the $4^N$ Bloch tensor coefficients $R_{\cal S}= R_{\mu_1,\ldots,\mu_N}$  
are real-valued (because of $\rho=\rho^\dagger$)
and satisfy $R_{0,\ldots, 0}=1$ (because of state normalization).  
Using a standard tensor product representation in the computational basis ($\sigma_j=0,1$), 
a pure state is parametrized by $2^N$ complex numbers $C_{\sigma_1,\ldots,\sigma_N}$ subject to normalization,
\begin{equation}\label{tensorprodrep}
|\Psi\rangle=\sum_{\sigma=\{\sigma_1,\ldots,\sigma_N\}} C_{\sigma} |\sigma\rangle,
\qquad |\sigma\rangle=|\sigma_1,\sigma_2,\ldots,\sigma_N\rangle,
\end{equation}
where the Bloch tensor follows as
\begin{equation}\label{Blochconversion}
    R_{\cal S} = \sum_{\sigma,\sigma'}     C^\ast_{\sigma}C_{\sigma'}   
    \prod_{j=1}^N \langle \sigma_j| \sigma_j^{\mu_j} |\sigma'_j\rangle.
\end{equation}
For instance, for $|\Psi\rangle=|00\cdots 0\rangle$, one finds $R_{\cal S}=
\prod_{j=1}^N (\delta_{\mu_j,0}+\delta_{\mu_j,3})$.

The stochastic master equation \eqref{drho2} then determines the measurement-conditioned  change of the 
Bloch tensor in one time step,
\begin{equation}\label{drs}
dR_{\cal S} =  R_{\cal S}(t+\delta t)- R_{\cal S}(t)={\rm Tr}(d\rho\, {\cal S}),
\end{equation}
see Eq.~\eqref{dRexpl} for an explicit expression.
Averaging over measurement outcomes, the Lindbladian form of Eq.~\eqref{ddrho2} gives
\begin{eqnarray}\nonumber
\overline{dR_{\cal S}} &=& - 2 \delta t\sum_{m=n,n+1} \sum_{\alpha\ne \alpha_m} \Bigl (  s_mJ_m\, \delta_{\beta_m,z}  \sum_{\alpha'}  \varepsilon_{\alpha_m\alpha\alpha'} + \\
    && \qquad+\, \Gamma_m \delta_{\beta_m,\perp} \Bigr)  \delta_{\mu_m,\alpha} R^{}_{{\cal S}} , \label{davR}
\end{eqnarray}
with the standard Levi-Civita symbol $\varepsilon$. 

By taking a partial trace over the Hilbert space pertaining to one (or several) qubits, one obtains the RDM  describing 
the state of the remaining qubits \cite{Nielsen2000,Zeng2019}.  In many cases, this state is mixed. 
For the RDM $\rho^{(r)}_{{\cal M}}$, one selects an ordered subset of $r<N$ qubits, 
${\cal M}=\{j_1,j_2,\ldots, j_r\}$ with $1\le j_1<\cdots<j_r\le N$, and partially traces over the $N-r$ remaining other qubits.  
With the complementary set  ${\cal M}_c$ corresponding to the traced-out qubits, we have
 $\rho^{(r)}_{\cal M}= {\rm Tr}_{{\cal M}_c}^{} \rho$.
To proceed, we again use Pauli string operators ${\cal S}_{\cal M}$ as in Eq.~\eqref{string} 
but keeping only operators $\sigma_{j}^{\mu_j}$ with $j\in {\cal M}$, i.e., 
operators for traced-out qubits are left out when
constructing the length-$r$ string operator ${\cal S}_{\cal M}$ from the original length-$N$
string operator ${\cal S}$. 
Since Pauli matrices are traceless, Eq.~\eqref{Bloch} then yields 
\begin{equation}\label{Blochk}
    \rho^{(r)}_{\cal M} = \frac{1}{2^r} \sum_{{\cal S}_{\cal M}} 
    R^{(r)}_{{\cal S}_{\cal M}} {\cal S}_{\cal M},
\end{equation}
where the rank-$r$ Bloch tensor $R^{(r)}_{{\cal S}_{\cal M}}$ 
follows from the Bloch tensor $R_{\cal S}$ for the pure state $\rho=|\Psi\rangle\langle\Psi|$ 
  by simply putting $\mu_j=0$ for all traced-out qubits ($j\in {\cal M}_c$).
  
Let us briefly give a few examples. For $N=1$, Eq.~\eqref{Bloch} reproduces the standard Bloch vector representation of 
a single-qubit state, $\rho=\frac12 \left( \sigma^0 + \mathbf{R}\cdot {\bm \sigma} \right),$
with the Bloch vector $\mathbf{R}=(R_1,R_2,R_3)^T$ and ${\bm \sigma}=(\sigma^x,\sigma^y,\sigma^z)$. 
For $N=2$, Eq.~\eqref{Bloch} involves the $4\times 4$ Bloch 
matrix $R_{\mu_1,\mu_2}$ \cite{Gamel2016}.
The entries $R_{\alpha_1,0}$ and $R_{0,\alpha_2}$ with $\alpha_j\in\{1,2,3\}$  determine
the Bloch vectors $\mathbf{R}_1$ and $\mathbf{R}_2$ for the single-qubit RDMs 
$\rho^{(1)}_1$ and $\rho^{(1)}_2$ corresponding to the first and second 
qubit, respectively.  The remaining entries encode the correlator \eqref{spincor},  $R_{\alpha_1,\alpha_2}=Q_{1,2}^{\alpha_1,\alpha_2}$. For $N=3$, the state is represented by $R_{\mu_1,\mu_2,\mu_3}$.
The Bloch vector $\mathbf{R}_{1}$ determining the single-qubit RDM for the first qubit, $\rho_1^{(1)}={\rm Tr}_{2,3}\rho$, follows as ${\bf R}_1=(R_{1,0,0}, R_{2,0,0}, R_{3,0,0})^T$, and likewise for other qubits.
Similarly, there are three two-qubit RDMs, $\left\{ \rho^{(2)}_{12}, \rho_{23}^{(2)}, \rho_{13}^{(2)} \right\}$. 
For instance,  $\rho_{12}^{(2)}$ --- the mixed state of qubits 1 and 2 after tracing over qubit 3 --- 
is represented by $R^{(2)}_{\mu_1,\mu_2}=R_{\mu_1,\mu_2,0}$ according to Eq.~\eqref{Blochk}.   The matrix elements $R^{(2)}_{\alpha_1,0}$ and $R^{(2)}_{0,\alpha_2}$ encode the Bloch vectors $\mathbf{R}_1$ and $\mathbf{R}_2$, while
$R^{(2)}_{\alpha_1,\alpha_2}=Q_{1,2}^{\alpha_1,\alpha_2}$.
 
In what follows, the Bloch tensor representation of the 
target state  $\rho^f=|\Psi_f\rangle\langle\Psi_f|$  is denoted by $R^f_{\cal S}$,
which also determines the Bloch tensors $R^{(r) f}_{{\cal S}_{\cal M}}$ for 
the  corresponding $r$-qubit RDMs $\rho^{(r) f}_{\cal M}$. 

\subsection{Active feedback and cost function}\label{sec2c}

We next discuss the active decision making part of the protocol, where the steering parameters $K=(K_n,K_{n+1})$ 
used in the next time step are selected. In our protocol, this decision is
based on optimizing the expected (i.e., averaged over measurement outcomes) 
change of a cost function, $\overline{dC(K)}$, with
respect to the steering parameters, see Eq.~\eqref{dck}.  
If several $K$ form a set with the same optimal average cost function gain, we choose 
$K$  according to a uniform random distribution from this set.  

We note in passing that interesting variants of our protocol could be based on other strategies.  For instance, 
one could optimize an entanglement measure such as the entanglement entropy in order to steer
states towards a class of highly entangled target states.
Grouping $N$ qubits into two subsets $A$ and $B$, and defining the RDM $\rho_A={\rm Tr}_B \rho(t)$,
the entanglement entropy is given by \cite{Nielsen2000}
\begin{equation}\label{EE}
    S(t) = - {\rm Tr}_A \left(\rho_A \, \ln\rho_A \right)  . 
\end{equation}
The measurement-conditioned change, $dS=S(t+\delta t)-S(t)=- {\rm Tr}_A \left(d\rho_A  \ln\rho_A \right)$,
then has a nonlinear dependence on $\rho$.  While the use of entanglement measures 
for active feedback protocols is beyond the scope of this work,  we show results for $S(t)$ in Sec.~\ref{sec3} 
in order to illustrate how entanglement is generated (or removed) in the active steering process.

For state preparation, one cost function candidate is the fidelity-based
cost function $C_N=1-{\rm Tr}(\rho\rho^f)$ in Eq.~\eqref{cost}. For a given steering parameter choice $K$, 
the expected cost function change is given by
\begin{equation}\label{dCNK}
    \overline{dC_N(K)}  =- {\rm Tr}(\overline{d\rho} \,\rho^f)=-\frac{1}{2^N} \sum_{\cal S} \overline{dR_{\cal S}} \,R_{\cal S}^f,
\end{equation}
with $\overline{d\rho}$ in Eq.~\eqref{ddrho2} and $\overline{dR_{\cal S}}$ in Eq.~\eqref{davR}.
Interestingly, the same quantity can equivalently be expressed in terms of weak values 
 \cite{Aharonov1988}. 

\subsubsection{Weak Value representation}

The weak value representation of Eq.~\eqref{dCNK} follows by writing $\overline{dC_N(K)} = - \langle \Psi_f| \overline{d\rho} |\Psi_f\rangle$ and using Eq.~\eqref{ddrho2}.  Assuming
that the time-evolving state $|\Psi\rangle$ is \emph{not} orthogonal to the target state $|\Psi_f\rangle$ (i.e., $C_N<1$), we find 
\begin{eqnarray}\nonumber
  \frac{\overline{dC_N(K)}}{1-C_N}  &=&  \delta t\sum_{m=n,n+1} \Bigl
  [  -2s_mJ_m \delta_{\beta_m,z}\, {\rm Im}(W_{m,\alpha_m}) \\
 \label{WV}  &+& \Gamma_m\delta_{\beta_m,\perp} \left(1- | W_{m,\alpha_m} |^2\right)  \Bigr],
\end{eqnarray}
with complex-valued weak values \cite{Aharonov1988} pertaining to single-qubit Pauli operators,
\begin{equation}\label{WV2}
  W_{m,\alpha_m} = \frac{\langle \Psi_f| \sigma_m^{\alpha_m} |\Psi\rangle}{\langle\Psi_f|\Psi\rangle}.
\end{equation}

If the target state has been reached, $|\Psi\rangle=e^{i\gamma}|\Psi_f\rangle$, 
all $W_{m,\alpha_m}$ are real-valued (because $\sigma_m^{\alpha_m}$ is Hermitian) and
satisfy $|W_{m,\alpha_m}|\le 1$.  Evidently, Eq.~\eqref{WV} then tells us that one cannot 
find steering parameters that could further improve the average cost function, precisely as expected at convergence. If 
convergence has not yet been reached, however, successful steering requires that at least one steering 
configuration $K$ exists such that the cost function can still be improved on average, $\overline{dC_N(K)}<0$. 
According to Eq.~\eqref{WV}, if ${\rm Im} (W_{m,\alpha_m}) \ne 0$ holds for at least
one choice of ($m,\alpha_m$), 
one can satisfy $\overline{dC_N(K)}<0$ for the global fidelity cost function by choosing $\beta_m=z$ 
and $s_m={\rm sgn}[{\rm Im}(W_{m,\alpha_m})].$  Since
for nearly orthogonal states $|\Psi\rangle$ and $|\Psi_f\rangle$,
weak value matrix elements tend to be large  (``weak value amplification'' \cite{Aharonov1988}),
 Eq.~\eqref{WV} suggests that a particularly large improvement of the cost function can arise 
 in this case.

However, it may happen that \emph{all} possible weak values in Eq.~\eqref{WV2} are real-valued with $|W_{m,\alpha_m}|\le 1$ before convergence has been reached, 
i.e., for $C_N>0$.  Equation \eqref{WV} then predicts $\overline{dC_N(K)}\ge 0$ for all possible
$K$, and the state trajectory gets stuck in a state where  
the average cost function cannot be lowered anymore by any steering operator.  In what follows, 
we refer to such states as ``trapped states.'' 
Active steering protocols based on the global fidelity cost function $C_N$ 
then encounter a prohibitive roadblock.  
Remarkably, trapped states are abundant already 
for just $N=2$ qubits, and they tend to proliferate with increasing $N$. 
For instance, consider a specific $N=2$ Bell target state 
\begin{equation}\label{EPR}
    |\Psi_f\rangle=|{\rm Bell}\rangle=(|00\rangle+|11\rangle)/\sqrt2,
\end{equation}
where we can easily show that $|\Psi\rangle=|00\rangle$ is a trapped state. 
Indeed, for this example, we find $C_2=1/2$ and the weak values \eqref{WV2} are given by
$W_{m,\alpha_m}=\delta_{\alpha_m,z}$.  Hence Eq.~\eqref{WV} implies $\overline{dC_2(K)}\ge 0$ for all  $K$.  

In general, trapped states form a manifold in Hilbert space whose dimensionality 
depends on the target state.  Let us briefly illustrate this point 
for a two-parameter class of even-parity target states for $N=2$,  
\begin{equation}\label{Belltype}
|\Psi_f\rangle = u|00\rangle+ e^{i\theta} \sqrt{1-u^2} |11\rangle, 
\end{equation}
with real parameters $u\in[0,1]$ and $\theta\in[0,2\pi)$. 
Computing the weak values in Eq.~\eqref{WV2}, we find from Eq.~\eqref{WV} that 
for $u\ne \frac{1}{\sqrt2}$, the states 
\begin{equation}\label{unsteer}
    |\Psi\rangle = v|00\rangle + e^{i\theta} \sqrt{1-v^2} |11\rangle
\end{equation}
with arbitrary $v\in [0,1]$ span a one-dimensional trapped-state manifold.  
For $u= \frac{1}{\sqrt2}$, the dimensionality of the manifold can further increase because trapped states may now
also receive contributions from the odd-parity sector.

We conclude that the weak values in Eq.~\eqref{WV2}, evaluated along the time-evolving state trajectory, 
can provide useful hints about steering protocols. In particular, they are able to diagnose trapped-state manifolds 
which arise when all accessible weak values are real-valued with $|W_{m,\alpha_m}|\le 1$.

We note that there is another mechanism which can also invalidate active steering based on the global 
fidelity cost function.  This mechanism
arises from the orthogonality catastrophe problem discussed in Sec.~\ref{sec1}.  
In fact, for orthogonal states $|\Psi\rangle$ and $|\Psi_f\rangle$, the weak value matrix elements in Eq.~\eqref{WV2} 
are ill-defined.  However, $\overline{dC_N(K)}$ in Eq.~\eqref{WV} can then be written in the form
\begin{equation}
    \overline{dC_N(K)}  = - \delta t\sum_{m=n,n+1} 
    \Gamma_m\delta_{\beta_m,\perp} \, | \langle\Psi_f|\sigma_{m}^{\alpha_m} |\Psi\rangle|^2 .
\end{equation}
If all matrix elements $\langle\Psi_f|\sigma_m^{\alpha_m}|\Psi\rangle$ of single Pauli operators vanish, we have $\overline{dC_N(K)}=0$ and the steering protocol gets stuck, i.e., $|\Psi\rangle$ is a trapped state. 
This case is typically encountered for orthogonal many-qubit states.

The above discussion shows that the averaged fidelity-based cost function \eqref{cost} is not able to guarantee
the success of active steering protocols because of the presence of trapped states. Such states are 
either diagnosed by weak values or arise because of the orthogonality catastrophe problem.   
In addition, the separation of $\overline{dC_N(K)}$ into single-qubit weak values, see Eq.~\eqref{WV}, 
indicates that entanglement structures are not captured.  This separation is 
a consequence of the linearity of $C_N$ in $\rho$ and of the uncorrelated Lindblad dynamics 
of $\overline{d\rho}$ in Eq.~\eqref{ddrho2}. We will now describe how one can resolve this apparent roadblock.
 
\subsubsection{Local cost function terms}

In order to enable successful steering under protocols with active feedback,
we next include local fidelity contributions in the cost function. 
To that end, we introduce basis-independent terms which enforce that all
RDMs formed from the state $\rho(t)$ will approach the 
respective RDMs for $\rho^f$. This strategy is able to resolve the 
trapped-state problem. 

To that end, let us define a local cost function term, $C_r(t)$, for $r$-qubit RDMs with $r<N$.
This term includes contributions from all ordered subsets ${\cal M}$ which 
can be constructed for $r$ qubits, see Sec.~\ref{sec2b}.
The number of such subsets is given by a binomial coefficient,
$N_r= \left( \begin{array}{c} N\\ r\end{array} \right)$.
Using the squared Frobenius norm (\emph{aka} relative purity) \cite{Nielsen2000} to quantify the
distance between the RDM and the corresponding target RDM, we arrive at
\begin{equation}\label{localcost}
    C_r (t) = \frac{1}{2N_r} \sum_{\cal M} {\rm Tr} \left( \rho^{(r)}_{\cal M}(t) -\rho^{(r) f}_{\cal M} \right)^2.
\end{equation}
This cost function penalizes $r$-qubit RDMs which deviate from the respective $r$-qubit RDM of the target state.  
We observe that for $r=N$, Eq.~\eqref{localcost} reduces to $C_N(t)$ in Eq.~\eqref{cost}.  
Using the Bloch tensor representation \eqref{Blochk}, we obtain
\begin{equation}
    C_r(t)= \frac{1}{2^{r+1} N_r} \sum_{\cal M} \sum_{{\cal S}_{\cal M}} \left
    (R_{{\cal S}_{\cal M}}^{(r)}(t)-R_{{\cal S}_{\cal M}}^{(r) f}\right)^2,
\end{equation}
where we recall that the Bloch tensors  $R^{(r)}_{{\cal S}_{\cal M}}(t)$ for RDMs follow from the Bloch tensor 
$R_{\cal S}(t)$ for the pure state  $\rho(t)=|\Psi(t)\rangle\langle\Psi(t)|$
by setting $\mu_j=0$ for all traced-out qubits, see Sec.~\ref{sec2b}.  
Given the time-evolving Bloch tensor $R_{\cal S}(t)$, it is then straightforward to
determine the cost functions $C_r(t)$.
It is worth mentioning that one could employ other distance measures 
instead of the squared Frobenius norm in Eq.~\eqref{localcost}, e.g.,
the max norm, the trace norm, or a Schatten $p$-norm \cite{Weidmann1980}. However, we expect that the corresponding active steering protocols perform with similar efficiency.

Importantly, all cost function terms $C_r(t)$ with $r=1,\ldots,N$ are non-negative and 
simultaneously minimized by $C_r=0$ when the time-evolving state converges to the target state.
This scenario is reminiscent of the parent Hamiltonian construction in Ref.~\cite{Ticozzi2012},
where passive steerability of a state was shown to be equivalent to 
the simultaneous minimization of all local contributions of a parent Hamiltonian.  
However, our cost-function based approach is more general and does not suffer from the 
restrictions on steerability discussed in Ref.~\cite{Ticozzi2012}.

We here employ the weighted-sum method \cite{Ehrgott2005}, that is, we use a single cost function $C(t)$ which
takes into account the $r$-qubit contribution $C_r(t)$ with probability weight $p_r$,  
\begin{equation}\label{totalcost}
    C(t) = \sum_{r=1}^N p_r C_r(t),\quad \sum_r p_r=1.
\end{equation}
This function is  minimized, with $C=0$, by the converged state, $|\Psi\rangle=e^{i\gamma}|\Psi_f\rangle$. 
The expected cost function change in one time step, $\overline{dC(K)}=\sum_r p_r\, \overline{dC_r(K)}$, see Eq.~\eqref{dck},
then follows from 
\begin{eqnarray}\nonumber
    \overline{dC_r(K)} &=& \frac{1}{2^r N_r} \sum_{{\cal M}, {\cal S}_{\cal M}} 
    \Biggl[  \left(R^{(r)}_{{\cal S}_{\cal M}} - R_{{\cal S}_{\cal M}}^{(r)\,f}
    \right) \overline{dR^{(r)}_{{\cal S}_{\cal M}}} \\ &+& \frac{ 
    \overline{dR^{(r)\, 2}_{{\cal S}_{\cal M}}}}{2}\label{davcr}
    \Biggr],
\end{eqnarray}   
where $\overline{dR^{(r)}_{{\cal S}_{\cal M}}}$ follows from Eq.~\eqref{davR}.  We note that for $r=N$, Eq.~\eqref{davcr} 
reduces to Eq.~\eqref{dCNK}. The term $\overline{dR^{(r)\,2}_{{\cal S}_{\cal M}}}$ is due to the nonlinear state dependence
of the cost functions in Eq.~\eqref{localcost}. We provide its explicit form in  Eq.~\eqref{squared}.  
In fact, $\overline{dC_{r<N}(K)}$ has a nonlinear dependence on $d\rho$ because 
${\rm Tr}(\rho_{\cal M}^{(r) \, 2})\le 1$ is possible for RDMs.  
(For a pure state,  ${\rm Tr}(\rho^2)=1$ instead implies that the cost function change is linear in $d\rho$.)
Thanks to this nonlinearity,  the disentangling character of the averaged Lindblad dynamics
in Eq.~\eqref{WV} can be avoided. 

In practice, the efficiency of the active steering protocol depends on the choice of the 
probability weights $p_r$ in Eq.~\eqref{totalcost} \cite{Ehrgott2005}. 
Unfortunately, we have not found a simple strategy for determining the optimal weights.
Our heuristic choices are described in Sec.~\ref{sec3}, but future work
may be able to achieve a better understanding of this important issue.

\subsection{Convergence properties} \label{sec2d}

We here address the convergence properties of the protocol discussed above.
Even though we do not have a mathematically rigorous proof valid for arbitrary number $N$ of system qubits, the numerical simulation results for $N\le 6$ in Sec.~\ref{sec3} 
suggest that the protocol does converge for general $N$. In fact, in our simulations,  
we never encountered a case where a predesignated target state was out of reach.  
In addition, the arguments in Sec.~\ref{sec2a} suggest 
that the scheme does converge since one can in principle realize arbitrary sequences
of single- and two-qubit gates.  
In this section, for the case $N=2$, we provide analytical arguments indicating that our active steering protocol converges to a predesignated target state of arbitrary form. 
For the Bell state, these arguments are rigorous. 
The results below also illustrate how the inclusion of local cost functions 
resolves the trapped-states problem. 

For $N=2$, it suffices to take into account only steering operators with $\beta\in\{x,z\}$ in Eq.~\eqref{Pauligate}.   We use the Bloch vectors ${\bf R}_{1,2}$, with components $R_1^{\alpha}=R_{\alpha,0}$ and $R_2^\alpha=R_{0,\alpha}$, see Sec.~\ref{sec2b}, in order to parameterize the time-evolving system state. Similarly, ${\bf R}_{1,2}^f$ with components $R_m^{\alpha\, f}$  describes the target state.  
The average local cost function change \eqref{davcr} for $N=2$ can be written as
\begin{widetext}
\begin{eqnarray}\nonumber  
\overline{dC_1} &=& \frac{\delta t}{2} \sum_{m=1,2} \left( - s_mJ_m \delta_{\beta_m,z}
    ( {\bf R}_m\times {\bf R}^f_{m} )^{\alpha_m} +  
    \sum_{\alpha\ne \alpha_m} \Gamma_m\delta_{\beta_m,x} R_m^{\alpha}\, R^{\alpha \, f}_m\right) -\delta t \delta_{\beta_1,x}\delta_{\beta_2,x} 
    \frac{ \Gamma_1\Gamma_2 (\Gamma_1+\Gamma_2)X_{\alpha_1,\alpha_2} }{(\Gamma_1+\Gamma_2)^2-4\Gamma_1\Gamma_2 R^2_{\alpha_1,\alpha_{2}}}  , \\\label{N2dC1}
&& X_{\alpha_1,\alpha_2} =  \frac12 (1-R_{\alpha_1,\alpha_2}^2) ({\bf R}_1^2+{\bf R}_2^2) -   (R_1^{\alpha_1})^2 -
 (R_2^{\alpha_2})^2 +2R_{\alpha_1,\alpha_2} R_1^{\alpha_1} R_2^{\alpha_2},
\end{eqnarray}
where we used $Q_{1,2}^{\alpha_1,\alpha_2}=R_{\alpha_1,\alpha_2}$, see Eq.~\eqref{spincor}.  
Instead of the weak value representation \eqref{WV}, it is here more convenient to express
 the average global fidelity change $\overline{dC_2}$ in terms of Bloch tensors as well.
From Eqs.~\eqref{dCNK} and \eqref{davR}, we then obtain the equivalent representation  
\begin{equation} \label{N2dC2}
    \overline{dC_2} = \frac{\delta t}{2} \sum_{m} \Bigl[ -s_m J_m \delta_{\beta_m,z}
    \bigl\{  ( {\bf R}_m\times {\bf R}^f_{m} )^{\alpha_m} + \sum_{\alpha'} 
    ( {\bf S}_{m;\alpha'} \times {\bf S}^f_{m;\alpha'})^{\alpha_m} \bigr\} + \Gamma_m\delta_{\beta_m,x} 
    \sum_{\alpha\ne \alpha_m} \bigl\{ R_m^{\alpha}\, R^{\alpha \, f}_m + \sum_{\alpha'}
    S_{m;\alpha'}^\alpha  S_{m;\alpha'}^{\alpha\, f} \bigr\}\Bigr].
\end{equation}
\end{widetext}
For given $m\in\{1,2\}$ and $\alpha'\in\{x,y,z\}$, the vector ${\bf S}_{m;\alpha'}$ is here defined by the components $S_{m=1;\alpha'}^\alpha=R_{\alpha,\alpha'}$ and
$S_{m=2;\alpha'}^\alpha=R_{\alpha',\alpha}$.  
With the probability weight $p_1$ in Eq.~\eqref{totalcost}, the
average total cost function change follows as $\overline{dC}=p_1 \overline{dC_1}+(1-p_1) \overline{dC_2}$.

Let us first study $|\Psi_f\rangle= |{\rm Bell}\rangle$ as target state, see Eq.~\eqref{EPR},
where the Bloch vectors vanish, ${\bf R}_{1,2}^f=0$.  (Analogous arguments can be given for the other Bell states.)
For $p_1=0$ in Eq.~\eqref{totalcost}, i.e., without the local fidelity term,
a subclass of trapped states is then given by Eq.~\eqref{unsteer} with $\theta=0$.
This class is parametrized by $0\le v\le 1$ with $v\ne \frac{1}{\sqrt2}$.
Importantly, it is always possible to identify at least one steering operator such that 
$\overline{dC_1}<0$ for these states.  As  a consequence, the active steering protocol 
will converge to the Bell state by allowing for a finite weight $p_1>0$.
To see this, we note that ${\bf R}_{1,2}^f=0$ implies that only the $X_{\alpha_1,\alpha_2}$-term 
in Eq.~\eqref{N2dC1} can contribute to $\overline{dC_1}$.  
Accordingly, one has to choose $\beta_1=\beta_2=x$, where we
can specify at least one steering parameter set $(\alpha_1,\alpha_2)$ such that $X_{\alpha_1,\alpha_2}>0$. 
Indeed, for the states in Eq.~\eqref{unsteer}, we find $R_m^{\alpha_m}=(2v^2-1)\delta_{\alpha_m,z}$
and  $R_{\alpha_1,\alpha_2}=\delta_{\alpha_1,\alpha_2}  \tilde Q_{\alpha_1}$, with
$\tilde Q_{x}=-\tilde Q_y=2v\sqrt{1-v^2}$ and $\tilde Q_z=1$.  From 
Eq.~\eqref{N2dC1}, we thus obtain $X_{\alpha_1,\alpha_2}=0$ for $\alpha_1=z$ or $\alpha_2=z$,  but $X_{\alpha,\alpha}=(2v^2-1)^4$ for $\alpha\ne z$ and $X_{x,y}=X_{y,x}=(2v^2-1)^2$.  
One can therefore always lower the total cost function \eqref{totalcost} until convergence has
been achieved.  This convergence proof for the Bell state also illustrates how 
the inclusion of local cost functions resolves the trapped-states
problem and thereby allows for successful active steering.

We observe numerically that the protocol also converges to a predesignated 
target state $|\Psi_f\rangle$ of arbitrary form, with non-vanishing Bloch vectors ${\bf R}_m^f$.
In order to rationalize this observation,
we first observe that the terms $\sim s_m J_m$ in 
Eqs.~\eqref{N2dC1} and \eqref{N2dC2} imply that one can always lower the cost function 
by choosing $\beta_m=z$ and a suitable sign $s_m$ unless the Bloch vectors
${\bf R}_m$ are parallel (or antiparallel) to ${\bf R}_m^f$, respectively.  
Similarly, the cost function in Eq.~\eqref{N2dC2} can be lowered by $\beta_m=z$
terms until the vectors ${\bf S}_{\alpha}\equiv {\bf S}_{2;\alpha}$ 
satisfy the relations
\begin{equation}\label{eq1}
  \sum_{\alpha}  {\bf S}_{\alpha}^{}\times {\bf S}_{\alpha}^f =0
\end{equation}
and (for all pairs $\alpha<\alpha'$) 
\begin{equation}\label{eq2}
    {\bf S}_{\alpha}^{}\cdot {\bf S}^f_{\alpha'} = {\bf S}_{\alpha'}^{}\cdot {\bf S}_{\alpha}^f. 
\end{equation}
In addition, purity of the state, i.e., ${\rm Tr}(\rho^2)={\rm Tr}(\rho^{f\, 2})=1$, implies the relation
\begin{equation}\label{purecond2}
   \sum_m (a_m^2-1)({\bf R}_m^f)^2+ \sum_\alpha [({\bf S}_{\alpha})^2-({\bf S}^f_{\alpha})^2 ]= 0,
\end{equation}
where we write ${\bf R}_m = a_m {\bf R}_m^f$.
We can express the solution to Eqs.~(\ref{eq1}) and (\ref{eq2}) as 
\begin{eqnarray}\nonumber
    {\bf S}_{x} &=& b_f {\bf S}^f_x + b_\times {\bf S}_y^f\times {\bf S}_z^f + b_{xy} {\bf S}_y^f + 
    b_{xz} {\bf S}^f_z ,\\   \label{blochv2}
    {\bf S}_{y} &=& b_f {\bf S}^f_y + b_\times {\bf S}_z^f\times {\bf S}_x^f + b_{xy} {\bf S}_x^f + 
    b_{yz} {\bf S}^f_z ,\\ \nonumber
    {\bf S}_{z} &=& b_f {\bf S}^f_z + b_\times {\bf S}_x^f\times {\bf S}_y^f + b_{xz} {\bf S}_x^f + 
    b_{yz} {\bf S}^f_y ,
\end{eqnarray}
where the real coefficients $(b_f,b_\times, b_{xy},b_{xz},b_{yz})$ must satisfy Eq.~(\ref{eq2}) and ${\bf S}_\alpha$ has to be of pure-state form \cite{Gamel2016}.  
We then examine the effects of $\beta_m=x$ steering operators. 
From the terms aside the $X$-term in Eq.~\eqref{N2dC1}, 
convergence requires $a_m\ge 0$ and for $\alpha\in \{x,y,z\}$ the relations 
\begin{equation}
\sum_{\alpha'\neq\alpha}{\bf S}_{\alpha'}^{}\cdot {\bf S}_{\alpha'}^f \ge 0 ,\quad \sum_{\alpha'} \left( {\bf S}_{\alpha'}^{}\cdot {\bf S}^f_{\alpha'}-
    S^\alpha_{\alpha'}  S^{\alpha \, f}_{\alpha'}\right) \ge 0.
\end{equation}
The $X$-term is important since it is the only term coupling the 
Bloch vectors and the vectors ${\bf S}_\alpha$.  
A sufficient condition for successful steering arises if 
the only solution for the above relations is given by $a_m=1$ and $b_f=1$, with $b_\times=b_{xy}=b_{xz}=b_{yz}=0$.
The requirement that ${\bf S}_\alpha$ must be of the pure-state form
(\ref{blochv2}), together with all the above restrictions, suggests that there is essentially no freedom for other
solutions besides ${\bf S}_\alpha={\bf S}^f_\alpha$.
While the above argument is not a convergence proof, it gives further evidence 
for the fact that our steering protocol does converge for all possible $N=2$ target states.\\

\section{Simulation results}\label{sec3}

In order to test the active steering protocol introduced above, we have
performed numerical simulations of the stochastic Schr\"odinger equation \eqref{SSE}.
We provide general remarks on our numerical approach in Sec.~\ref{sec3a}, followed
by a presentation of numerical results for the simplest case $N=2$ in Sec.~\ref{sec3b}.  We 
here mainly focus on the Bell state in Eq.~\eqref{EPR}.  In Sec.~\ref{sec3c}, we turn to the
highly entangled GHZ and W states for $N$ qubits
in Eq.~\eqref{GHZn}.  We begin with the case $N=3$, and then turn to larger values of $N$.

\subsection{General remarks} 
\label{sec3a}

All cost function terms in Eq.~\eqref{totalcost} as well as the expectation values for the 
corresponding changes in Eq.~\eqref{davcr} can be computed from the time-evolving
Bloch tensor $R_{\cal S}(t)$ which in turn follows from Eq.~\eqref{Blochconversion}.
For every time step,  $[N/2]$ non-overlapping pairs of adjacent qubits are steered, 
where we scan through all possible steering parameters $K$ for each pair in order to 
identify the optimal choice giving the largest expectation value of the cost function gain.  
The steering parameters are taken from the Pauli operators \eqref{Pauligate}.
Unless noted otherwise, the numerical results shown below were obtained by excluding detector operators of Pauli type $\beta_n=y$ in Eq.~\eqref{Pauligate}.  (However, such terms are helpful when preparing the W state or for product states.)
Once the optimal steering parameters have been identified, the quantum measurement 
is simulated by stochastically determining the 
measurement outcome $(\xi,\eta)$ according to the probabilities $P_{\xi,\eta}$ in Eq.~\eqref{Pdef2}.  Subsequently, the quantum state is updated using Eq.~\eqref{SSE} 
and the respective detector qubit pair is re-initialized in the product state $|00\rangle_d$ before the next iteration step is launched.

For a given state trajectory, starting from the initial state $|\Psi(0)\rangle=|00\cdots 0\rangle$, the steering protocol is terminated once the fidelity $F(t)$ in Eq.~\eqref{cost} exceeds a predefined threshold value $F^\ast$ above which the state is considered converged.  
The corresponding time $t$ defines the number of time steps $n_t=t/\delta t$ needed for
reaching convergence.  Of course, this number varies for different measurement-resolved state trajectories.
By collecting a histogram from $M\gg 1$ trajectory realizations, we can obtain a numerical estimate for the probability distribution of the step number $n_t$. This distribution depends on the 
chosen fidelity threshold $F^\ast$.
As shown below, we find that the distribution is quite different from a Gaussian distribution and of a typical asymmetric shape.  (We have tried to fit our numerical results to commonly used distributions but did not find  satisfactory agreement.) For a qualitative description of the simulation results, we characterize the distribution
by three indicators, namely (i) the median $N_s$, (ii) the mode $N_{m}$ corresponding to the maximum, and (iii) the half-width $\Delta N$ defined as the width of the histogram at half-maximum height. All three numbers depend on the fidelity threshold $F^\ast$. 

Apart from the statistics of the number of iteration cycles needed for achieving convergence, 
we also show results for the time dependence of the fidelity cost function $C_N(t)=1-F^2(t)$, see Eq.~\eqref{cost}, both on the level of individual state trajectories and for averages taken over many realizations. Similarly, we monitor the time dependence of the total cost 
function $C(t)$ in Eq.~\eqref{totalcost}.
Since no steering operations are applied anymore once the fidelity threshold has been passed,  the averaged cost functions $\overline{C_r(t)}$ can depend on the chosen value of $F^\ast$ at long times.

Throughout, we assume identical system-detector couplings, $J_n=1$, and choose $\delta t=0.2$ as 
elementary time step such that $\Gamma_n=0.2$.  We have checked that weak asymmetries in the $J_n$-couplings and/or moderate changes of $\delta t$ do not cause qualitative changes.  

\begin{figure}[t]
\includegraphics[width=\columnwidth]{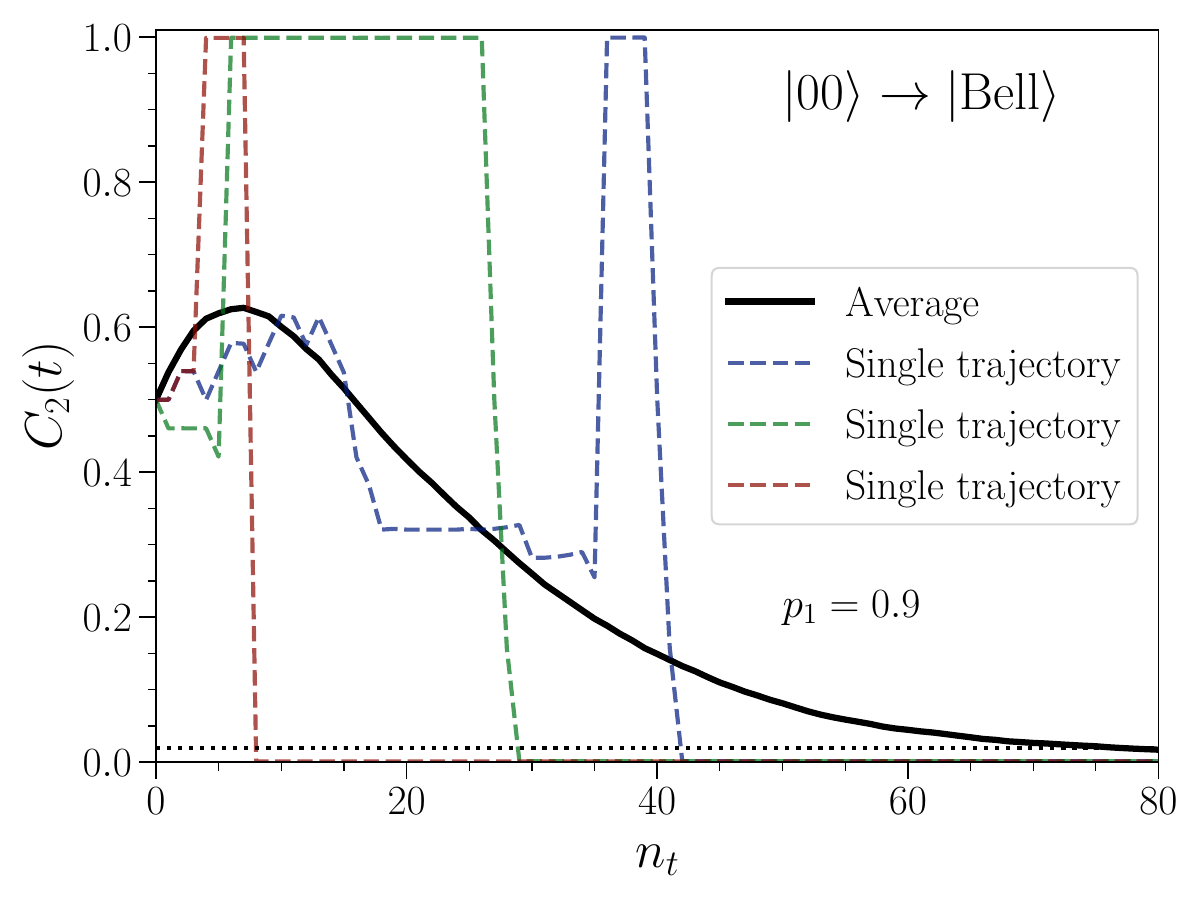}
\includegraphics[width=\columnwidth]{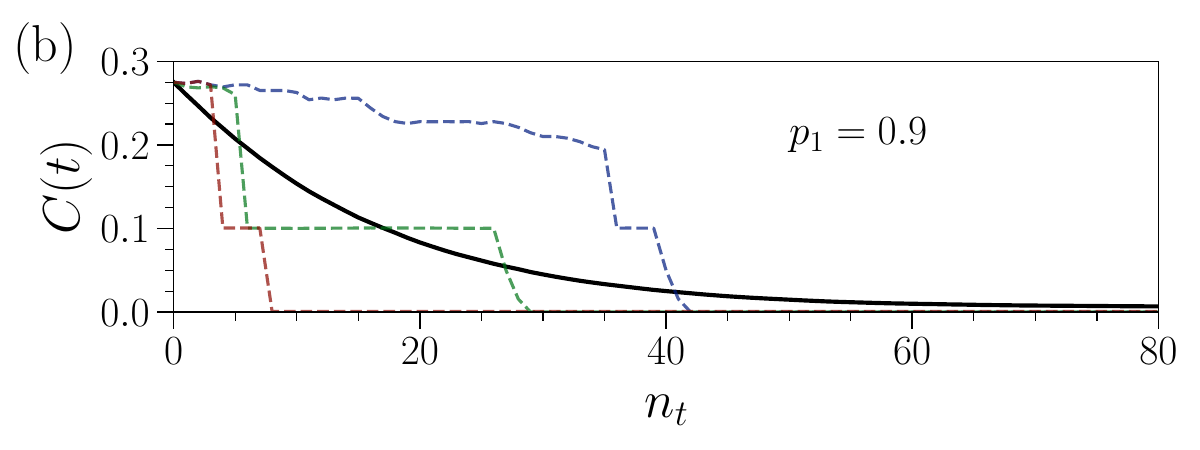}
\includegraphics[width=\columnwidth]{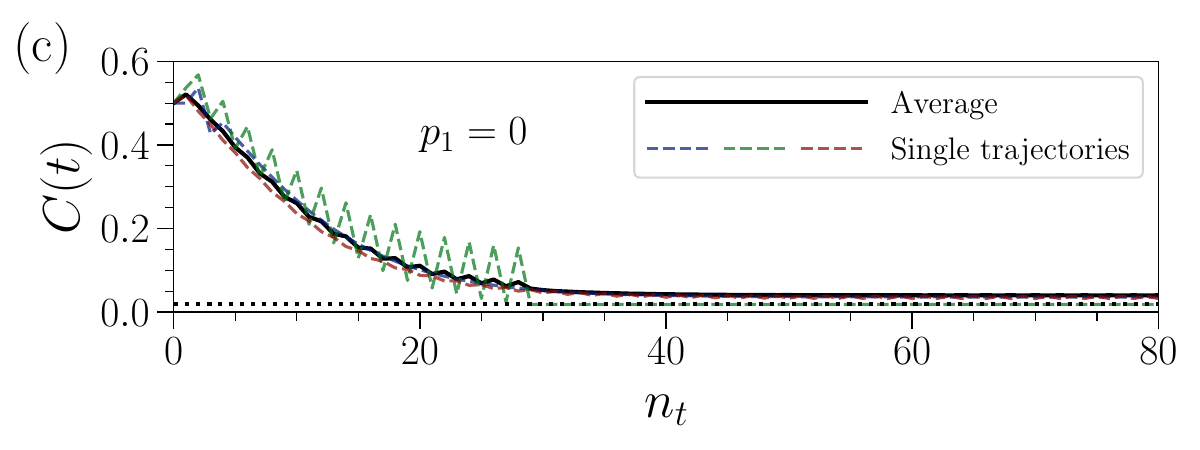}
\caption{ Active steering protocol for $N=2$ qubits and the target state $|\Psi_f\rangle=|{\rm Bell}\rangle$ in Eq.~\eqref{EPR}, using the total cost function $C(t)$ in Eq.~\eqref{totalcost} with  $p_1=1-p_2=0.9$ and target fidelity $F^\ast=99\%$. (a) Time evolution of the global fidelity cost function $C_2(t)=1-F^2(t)$ vs number of time steps $n_t=t/\delta t$. 
The colored dashed lines show three  measurement-resolved trajectories, the solid black curve is an average over $M=10^4$ runs. The dotted horizontal line indicates the fidelity threshold value.
(b) Corresponding results for the total cost function $C(t)$. 
(c) Corresponding results for $p_1=0$, where $C(t)=C_2(t)$ includes 
only the global fidelity cost function.  For the shown data, our simulations were modified to tolerate a tiny increase in $\overline{dC_2(K)}$ in order to still allow for 
active decision making.  Nonetheless, the fidelity threshold is only passed by very few 
state trajectories, while the great majority of trajectories fails to converge.
}
\label{fig3}
\end{figure}

\subsection{Bell state}\label{sec3b}

We start with the simplest case of $N=2$ qubits and study the active steering protocol for the 
target state $|\Psi_f\rangle=|{\rm Bell}\rangle$ in Eq.~\eqref{EPR}.  
Let us recall that all entangled $N=2$ states are equivalent under local operations and classical communication (LOCC) to $|{\rm Bell}\rangle$  \cite{Bennett1996}.
In fact, we have numerically tested for many examples that other $N=2$ target states can be 
reached with similar efficiency by our protocol.   For several measurement-resolved state trajectories and for an average taken over $M=10^4$ runs of the protocol,
Fig.~\ref{fig3}(a) shows the time evolution of $C_2(t)=1-F^2(t)$. 
Similarly, Fig.~\ref{fig3}(b) shows the total cost function $C(t)$ in Eq.~\eqref{totalcost}.
We here used the probability weights $p_1=0.9$ and $p_2=0.1$ for  the $r$-qubit cost function terms in Eq.~\eqref{totalcost}, but 
simulation results are qualitatively similar for all $p_1\agt 0.5$ (with $p_2=1-p_1$).  While the analytical argument given at the end of Sec.~\ref{sec2c} suggests that the trapped-states problem can be resolved for any $p_1>0$, we find that in practice, for small $p_1$, the protocol becomes inefficient.

\begin{figure}[t]
\centering
\includegraphics[width=\columnwidth]{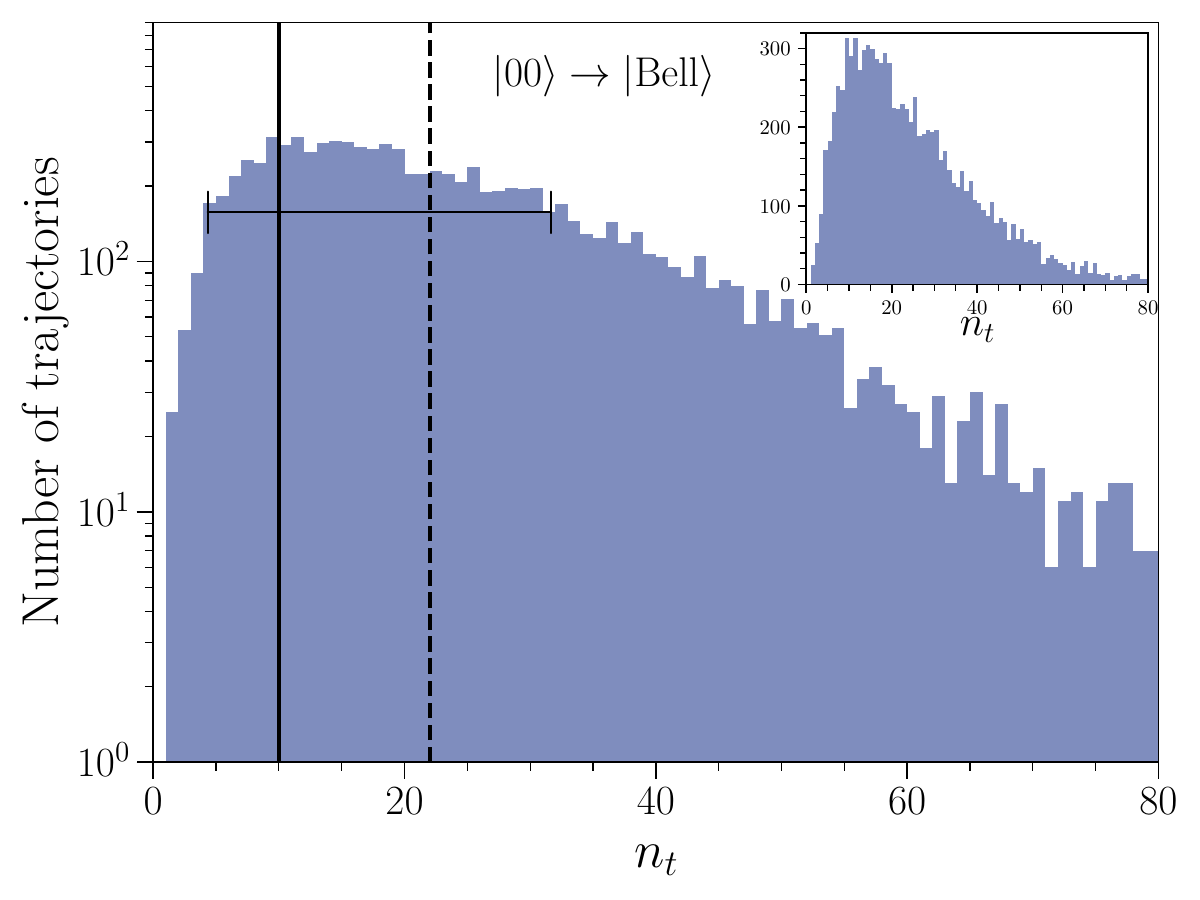}
\caption{ Histogram of the number of trajectories that have reached fidelity $F^\ast=99\%$ in $n_t=t/\delta t$ time steps, obtained from $M=10^4$ realizations of the $N=2$ protocol with $|\Psi_f\rangle=|{\rm Bell}\rangle$, see Fig.~\ref{fig3}.  Note the logarithmic scale for 
the vertical axis. The solid vertical line (horizontal bar) indicates the mode $N_m$ (half-width $\Delta N$), the dashed vertical line shows the median $N_s$. Inset:  Same results but with a linear scale for the vertical axis. }
\label{fig4}
\end{figure}

We observe that the individual state trajectories in Fig.~\ref{fig3} are characterized by strong fluctuations and large quantum-jump-induced steps in the global fidelity $F(t)$.    
Typically, the steering protocol $|\Psi(0)\rangle\to |\Psi_f\rangle$ does \emph{not} result in 
a monotonically increasing time dependence of the fidelity at the level of individual trajectories.   
Figure \ref{fig3}(a) shows that the protocol instead  first steers the system into one of the
other Bell states, which is orthogonal to the target state ($F=0$), but then just one additional quantum jump  is enough to bring the system to the target state.  
By allowing for a decrease of the global fidelity $F(t)$ at intermediate times, the protocol is thereby able to allow for fast and efficient active steering. Such state trajectories are easily found and implemented by our protocol 
because of the presence of the local cost function $C_1(t)$, see Sec.~\ref{sec2c}. It is worth emphasizing that, in our case, most state trajectories reach the target state with almost perfect fidelity, well above the fidelity threshold $F^\ast$ used in Fig.~\ref{fig3}.
We note in passing that the active decision framework of Ref.~\cite{Herasymenko2023} is consistent with this scenario. However, a crucial difference is that Ref.~\cite{Herasymenko2023} considers steering operators pertaining to passive steering protocols, where the active-decision policy only accelerates the steering. In the present protocol, active steering based on global and local cost functions instead allows one to ``steer the unsteerable'', i.e., to engineer passively unsteerable target states.

At the same time, the total cost function $C(t)$ is observed to monotonically decrease at the level of individual trajectories, see Fig.~\ref{fig3}(b), even though 
$C(t)$ could in principle increase after an ``unfavorable'' measurement outcome. In fact,
our protocol only enforces a decrease of the \emph{average} cost function $\overline{C(t)}$,  
which must be a monotonically decreasing function for $M\to \infty$. 
Similarly, after an initial transient behavior, the averaged squared global fidelity, $\overline{F^2(t)}=1-\overline{C_2(t)}$, increases monotonically, 
where we find an approximately exponential time dependence for approaching the target state, see
Fig.~\ref{fig3}(a).  Interestingly, the slope $\frac{d}{dt} \overline{F^2(t)}$ is  larger
for short-to-intermediate times than in the long-time limit:  states that are nearly orthogonal to the target state tend to be ``corrected'' more rapidly.  This observation is consistent with the weak value amplification mechanism discussed in Sec.~\ref{sec2c}.  

Next, Fig.~\ref{fig3}(c) shows the corresponding results obtained from the 
active steering protocol with $p_1=0$, where only the global fidelity cost function is used. 
In that case, steering is not successful because of the emergence of trapped states,
see Sec.~\ref{sec2c}.  In such cases, state trajectories get stuck before convergence has been reached, and the averaged  cost function saturates at some value above the target fidelity threshold.  While very few trajectories are still able to pass the fidelity
threshold, the scheme is inefficient and fails to converge.  

For the fidelity threshold $F^\ast=99\%$ and using $M=10^4$ realizations, Fig.~\ref{fig4} shows a histogram of the number of time steps $n_t$ needed for achieving convergence. 
Evidently, the corresponding distribution function of $n_t$ is asymmetric, quite broad, and 
rather different from a Gaussian distribution. This fact is particularly evident from the inset of 
Fig.~\ref{fig4}, where we use a linear scale for the vertical axis. 
Using the quantities introduced in Sec.~\ref{sec3a}
to characterize the distribution, we find the maximum (mode) $N_m= 10$, the
median $N_s= 22$, and the half-width $\Delta N=28$.
Importantly, there are many trajectories which end up in the 
target state after just a few cycles.
On the other hand, we also encounter rare trajectories which require an exceptionally large number of iterations for convergence.  

\begin{figure}[t]
\centering
\includegraphics[width=\columnwidth]{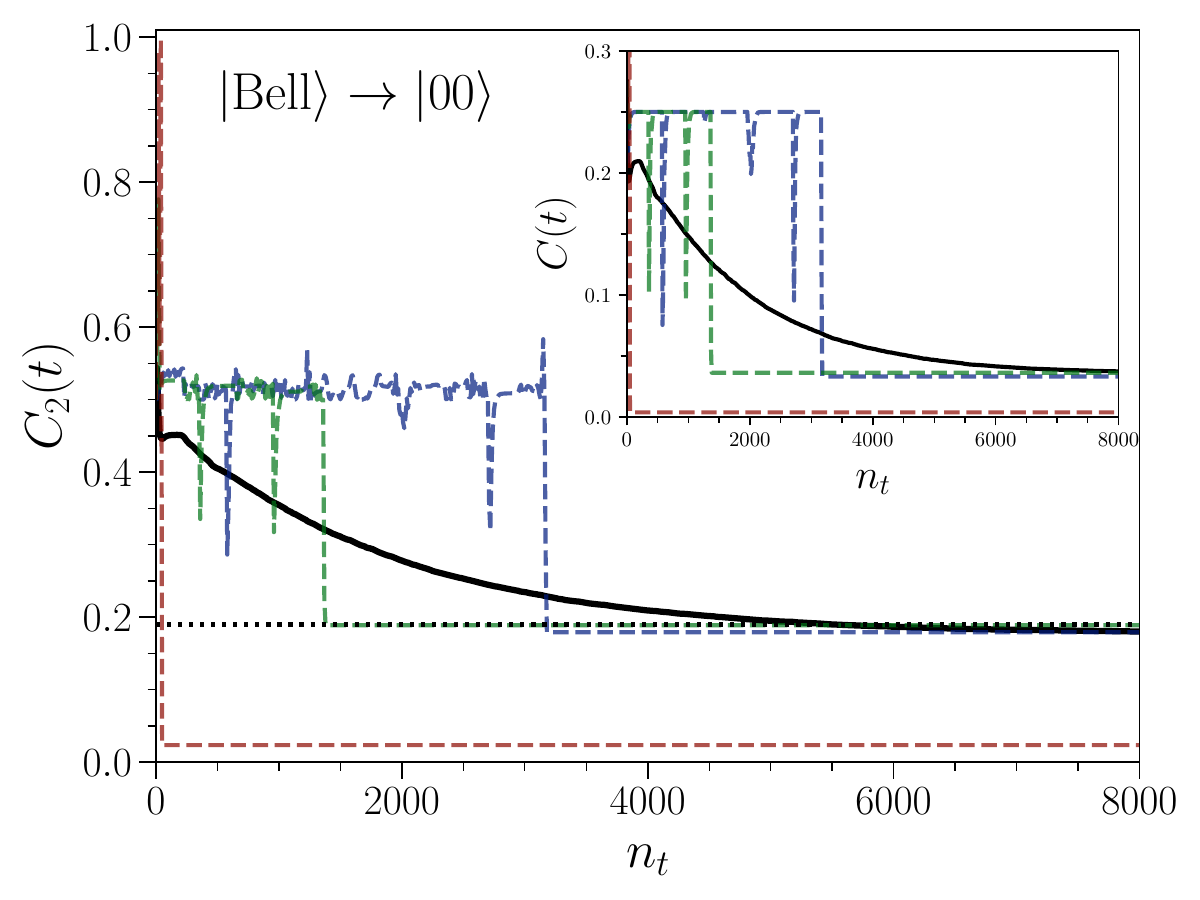}
\caption{ Reversed steering protocol for $N=2$ as in Fig.~\ref{fig3} but with interchanged initial and target states, $|\Psi(0)\rangle=|{\rm Bell}\rangle$ and $|\Psi_f\rangle =|00\rangle$, using $p_1=1-p_2=1$ in Eq.~\eqref{totalcost} and $F^\ast=90\%$. The steering operator set includes $\beta_m=y$ terms. 
Main panel: Time evolution of the global fidelity cost function $C_2(t)=1-F^2(t)$ vs number of time steps $n_t=t/\delta t$. Dashed colored curves show individual trajectories, the solid black curve is an average over $M=10^4$ realizations. 
The dotted horizontal line indicates the fidelity threshold $F^\ast$.  
Inset: Corresponding results for the total cost function $C(t)$. For $p_1=1$, we have $C=C_1$.
}
\label{fig5}
\end{figure}

In Fig.~\ref{fig5}, we study the time evolution of the global fidelity cost function $C_2(t)$ (main panel) and of the total cost function $C(t)$ (inset) if the protocol is run in
the opposite direction: we here start from $|\Psi(0)\rangle=|{\rm Bell}\rangle$ and steer towards $|\Psi_f\rangle=|00\rangle$. This seemingly trivial task, where entanglement needs to be removed from the system, is much more difficult to realize than the case in Fig.~\ref{fig3} using the present formulation of the protocol.
In fact, entanglement swapping as employed in our protocol gives efficient ways to inject entanglement into the system, but the removal of entanglement is harder to achieve.  Even though we target only the 
relatively poor fidelity $F^\ast=90\%$ in Fig.~\ref{fig5}, convergence times are now much longer.

From the main panel in Fig.~\ref{fig5}, we observe that different measurement-resolved trajectories surpass the fidelity threshold at widely different times, but they typically do not reach the target state with (almost) perfect fidelity as in Fig.~\ref{fig3}.  One of the shown trajectories 
passes the fidelity threshold after just 49 steps.  Since a finite fraction of all trajectories show such a behavior, we obtain a step-like initial decrease of the averaged fidelity cost function $\overline{C_2(t)}$
and of the averaged total cost function $\overline{C(t)}$ at very short times. 
While individual trajectories sometimes show a sudden increase of $C(t)$  
due to unfavorable measurement outcomes, the average $\overline{C(t)}$ 
decreases monotonically except for a shallow minimum at short times.  This
minimum is possible because of the finite number $M$ of runs used in computing the average.
Since one here typically ends up in states with imperfect fidelity, $F^\ast<F(t)<1$, where 
no steering operations are applied after the corresponding time anymore, 
$\overline{C(t)}$ and $\overline{C_2(t)}$ show a significant dependence on $F^\ast$ at long times.

The measurement-resolved trajectories in Fig.~\ref{fig5} show that an improvement in the global fidelity 
is often reversed again in the next step due to unfavorable measurement outcomes. Such a behavior
indicates that it is a nontrivial task for our protocol to remove entanglement from the system.
However, this task could easily be made highly efficient by adding occasional rounds of 
single-detector qubit measurements to the active steering protocol.  
Since our main interest is in the preparation of exotic highly entangled states,
however, we do not pursue this extension here. 
We conclude that, even though less efficient, the protocol is also able to actively steer in the backward direction. 

\begin{figure}[t]
\centering
\includegraphics[width=\columnwidth]{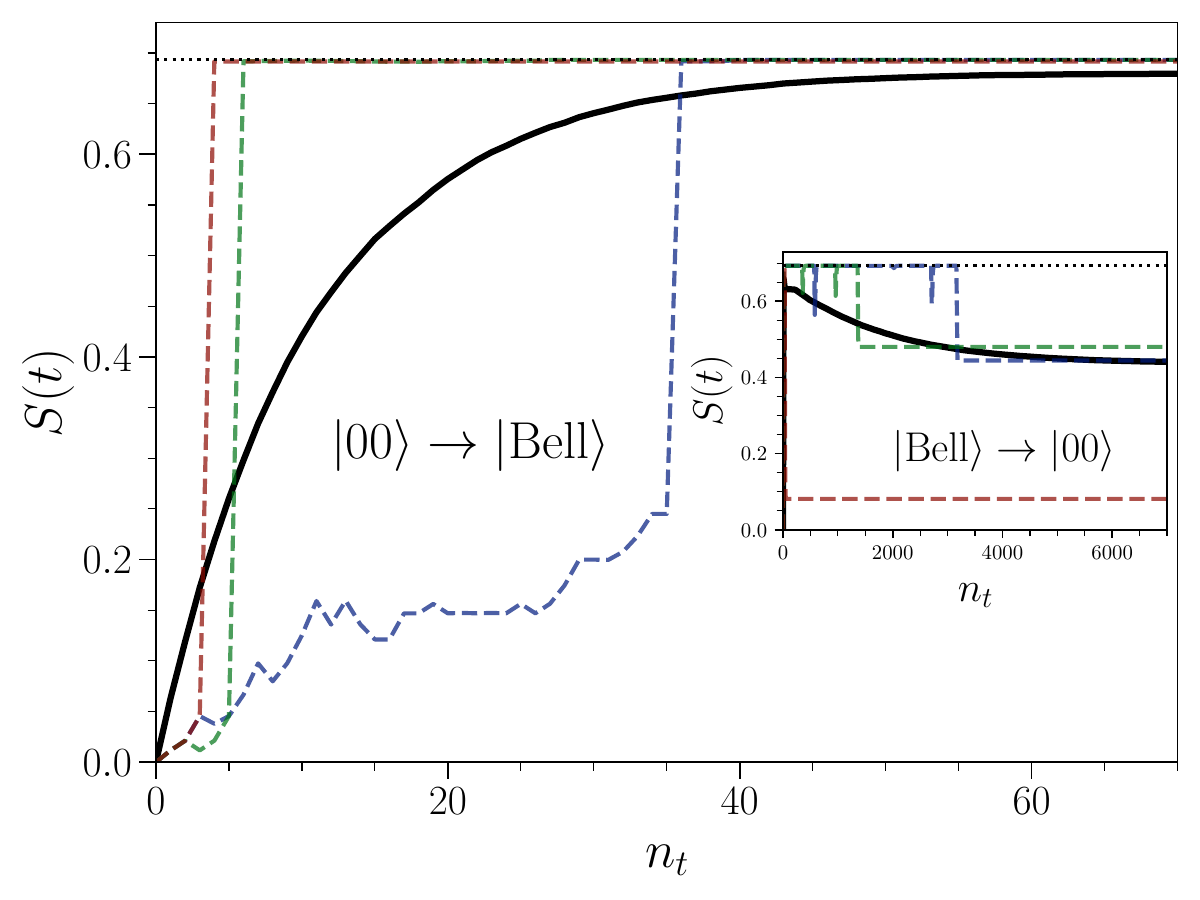}
\caption{Time evolution of the entanglement entropy $S(t)$ for $N=2$, see Eq.~\eqref{EE}.  
Main panel: $S(t)$  vs  $n_t=t/\delta t$ for $|\Psi_f\rangle=|{\rm Bell}\rangle$ for the same measurement-resolved  
trajectories as in Fig.~\ref{fig3}(a). The solid black curve is an average over $M=10^4$ runs. 
The dotted horizontal line indicates $S_{\rm max}=\ln 2$. Inset: Same but for the reversed protocol in Fig.~\ref{fig5}, where the low fidelity $F^\ast=90\%$ implies that $S(t)$ saturates at a relatively large value instead of approaching zero.  }
\label{fig6}
\end{figure}

The time evolution of the entanglement entropy $S(t)$ in Eq.~\eqref{EE} during the above $N=2$
steering protocols is shown in Fig.~\ref{fig6}. For the direction 
$|00\rangle \to |{\rm Bell}\rangle$, the main panel of Fig.~\ref{fig6} shows $S(t)$ 
for the same trajectories as in Fig.~\ref{fig3}(a). 
We observe that entanglement is quickly built up and
one approaches (at long times) the maximal value $S=\ln 2$ expected for $|{\rm Bell}\rangle$, 
see Refs.~\cite{Turkeshi2021,Sang2023} for related results.  In fact, since the trajectory typically cycles through other Bell states before reaching the final target state $|{\rm Bell}\rangle$, see Fig.~\ref{fig3}(a), the convergence 
of $\overline{S(t)}$ toward $S=\ln 2$ at long times is faster than the corresponding convergence of the fidelity cost function $\overline{C_2(t)}$ shown in Fig.~\ref{fig3}(a).
In the inset of Fig.~\ref{fig6}, using the trajectories in Fig.~\ref{fig5} 
for the reversed steering direction  $|{\rm Bell}\rangle\to |00\rangle$, we illustrate that 
 entanglement can also be removed from the system. 

 \begin{figure}[t]
\centering
\includegraphics[width=\columnwidth]{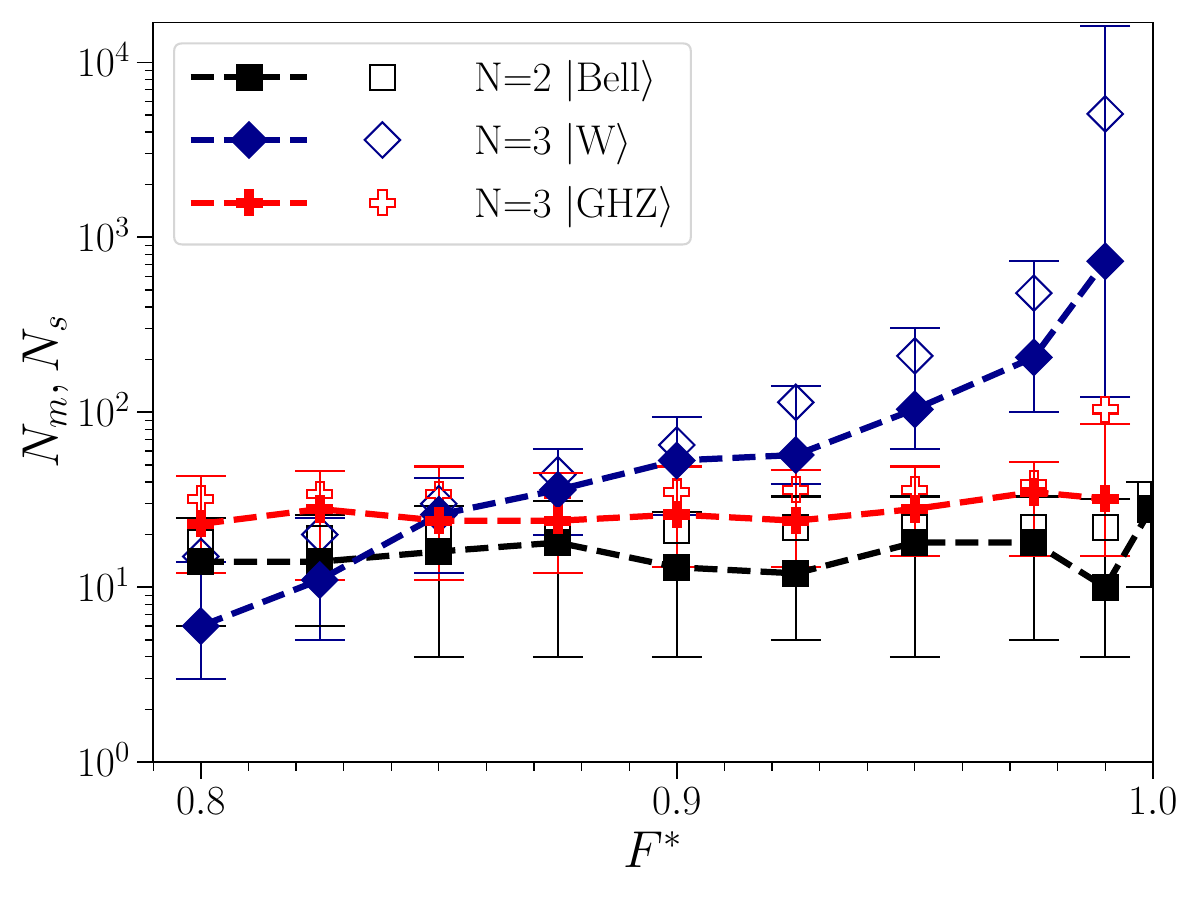}
\caption{Mode $N_m$ (filled symbols), median $N_s$ (open symbols), and half-width (vertical bars) vs target state fidelity $F^\ast$ for active steering to the $N=2$ target state 
$|{\rm Bell}\rangle$ and for the maximally entangled $N=3$ states $|{\rm GHZ}\rangle$ and 
$|{\rm W}\rangle$. 
The shown results have been obtained from $M=10^4$ trajectories. Note the logarithmic scale for the vertical axis. Dashed lines are a guide to the eye only.   
For the W state, steering operators with $\beta=y$ in Eq.~\eqref{Pauligate} have been included.
}
\label{fig7}
\end{figure}

In Fig.~\ref{fig7}, for several target states with $N=2$ and $N=3$, 
we show the mode $N_m$, the median $N_s$,  and the half-width $\Delta N$ 
 as  functions of the target fidelity $F^\ast$. These 
three numbers characterize the distribution function of the step number $n_t$, see Sec.~\ref{sec3a}, which in turn is estimated by collecting a histogram.  
Figure~\ref{fig7} indicates that the $F^\ast$-dependence of $N_m$ and $N_s$  is very weak for 
the  $|{\rm Bell}\rangle$ and $|{\rm GHZ}\rangle$ states, while it 
is approximately exponential for $|{\rm W}\rangle.$
We note that a steeper increase in $N_s(F^\ast)$ is observed for the $N=3$ states
in Fig.~\ref{fig7} in the limit $F^\ast\to 1$.  This increase 
can be rationalized by recalling that our termination policy, 
where one ceases to apply steering operations once a fluctuating state trajectory has 
passed the fidelity threshold, also affects the long-time limit of averaged cost functions. 
As a result, an exceptionally large number of steps is needed on average 
for reaching convergence if the target fidelity is very close to $F^\ast=1$.

\subsection{W and GHZ states} \label{sec3c}

 \begin{figure}
\centering
\includegraphics[width=\columnwidth]{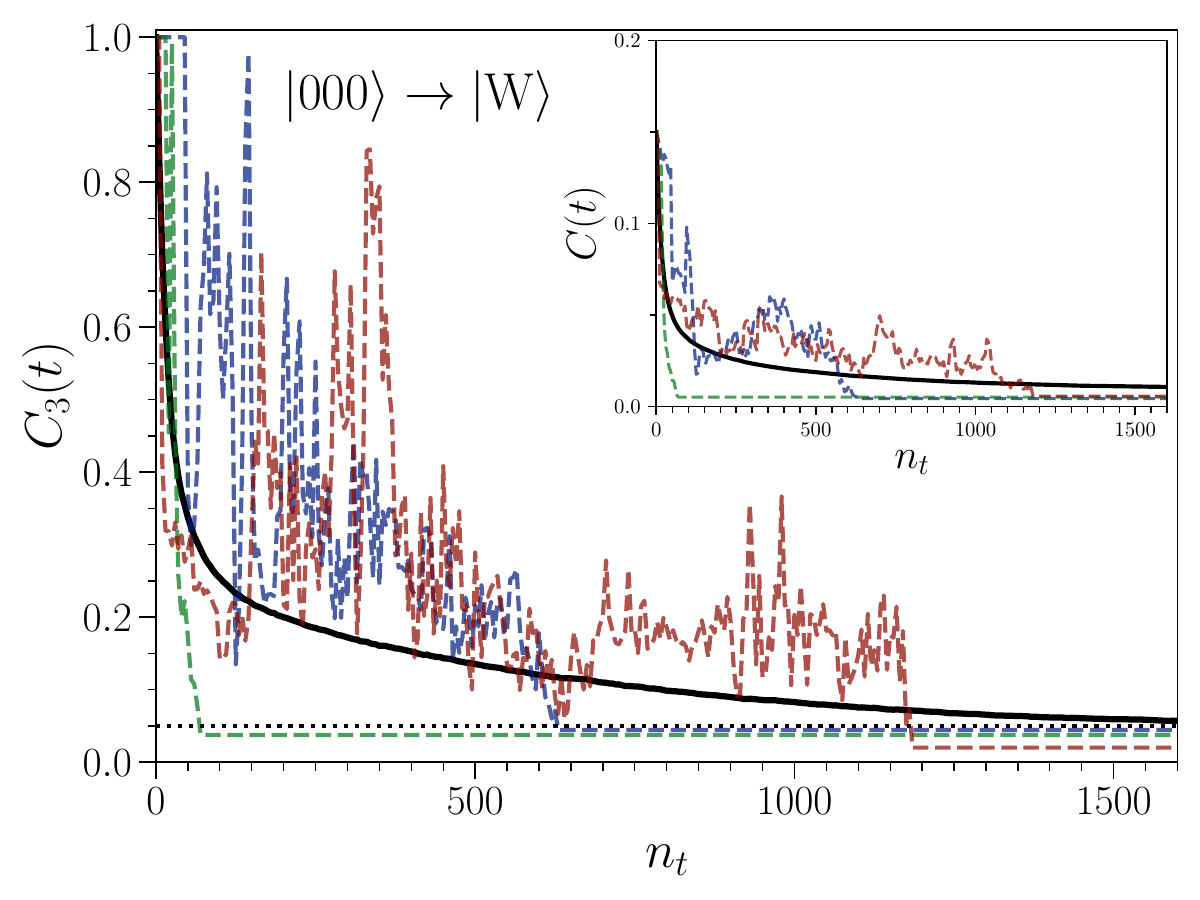}
\caption{Active steering protocol for $N=3$ qubits and the target state $|\Psi_f\rangle=|{\rm W}\rangle$ in Eq.~\eqref{GHZn}, with target fidelity $F^\ast=97.5\%$. We  use the total
cost function $C(t)$ in Eq.~\eqref{totalcost} with $p_1=0.9$, $p_2=0.09$ and $p_3=1-p_1-p_2=0.01$,
and include $\beta=y$ steering operators. 
Main panel: Global fidelity cost function $C_3(t)=1-F^2(t)$ vs number of time steps $n_t=t/\delta t$.  The dashed colored curves show three measurement-resolved  
trajectories, the solid black curve is an average over $M=10^4$ runs. 
The dotted horizontal line corresponds to the fidelity threshold.
Inset: Corresponding results for $C(t)$.
}
\label{fig8}
\end{figure}

\begin{figure}
\centering
\includegraphics[width=\columnwidth]{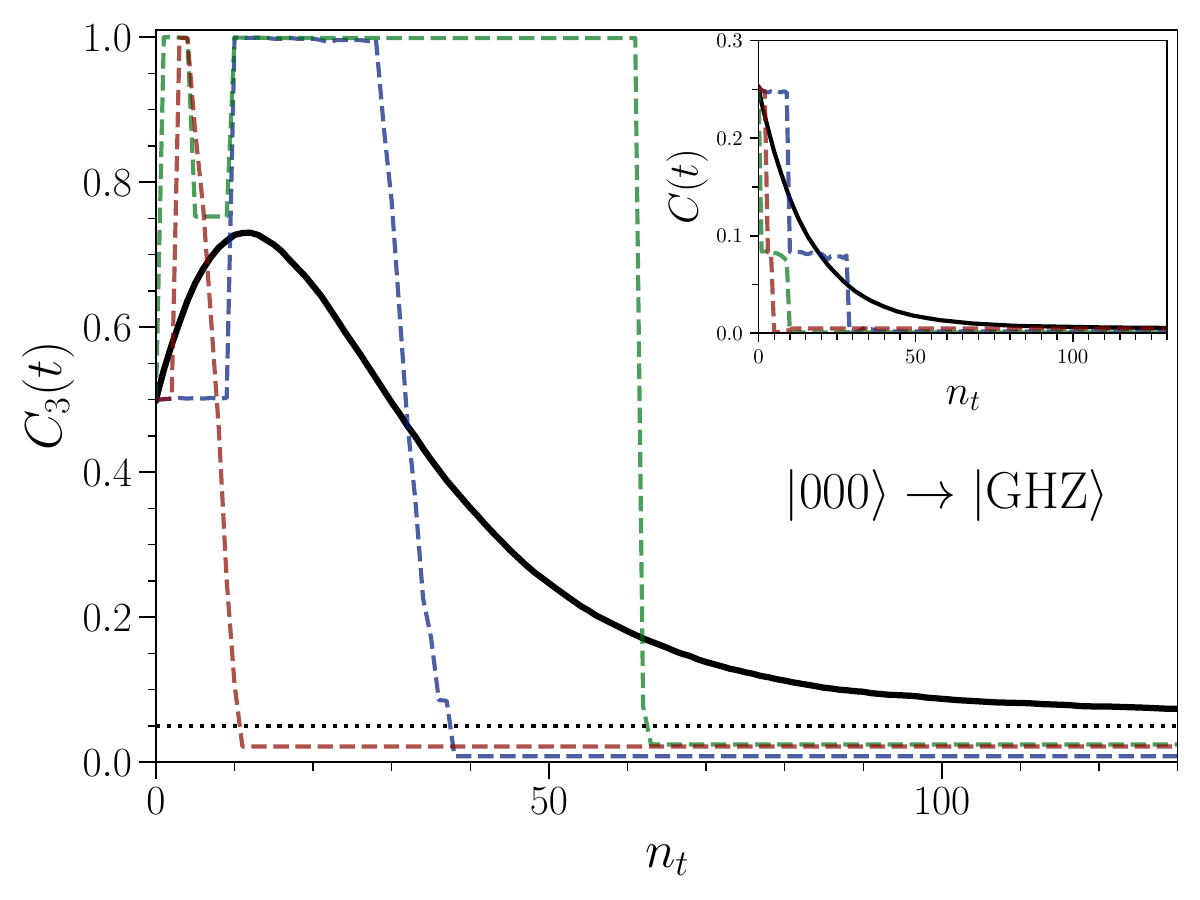}
\caption{Same as Fig.~\ref{fig8} but for the target state $|{\rm GHZ}\rangle$, again for
$F^\ast=97.5\%$ and with the same weights $p_r$.
Main panel: Global fidelity cost function $C_3(t)=1-F^2(t)$ vs  $n_t=t/\delta t$, 
where dashed colored curves correspond to individual trajectories and the black curve 
is an average over $M=10^4$ runs.  The dotted horizontal line corresponds to the fidelity
threshold. Inset: Corresponding results for the total cost function $C(t)$.}
\label{fig9}
\end{figure}

We next turn to the case $N=3$, where we consider active steering from $|\Psi(0)\rangle=|000\rangle$ to either
$|\Psi_f\rangle=|{\rm GHZ}\rangle$ or $|\Psi_f\rangle=|{\rm W}\rangle$ in Eq.~\eqref{GHZn}.
These two states represent different types of maximal tripartite entanglement \cite{Dur2000}.  However, we have 
numerically checked that the steering protocol performs with similar efficiency for many other $N=3$ target states, e.g., with additional phase factors in Eq.~\eqref{GHZn}.   
For all $N=3$ results shown here, we have used  the probability weights $p_1=0.9$, $p_2=0.09$, and $p_3=1-p_1-p_2=0.01$ for the total cost function in Eq.~\eqref{totalcost}.
Our protocol is then capable of finding both target states, but one needs a larger number of steps  than for $N=2$. 

\begin{figure}
\centering
\includegraphics[width=\columnwidth]{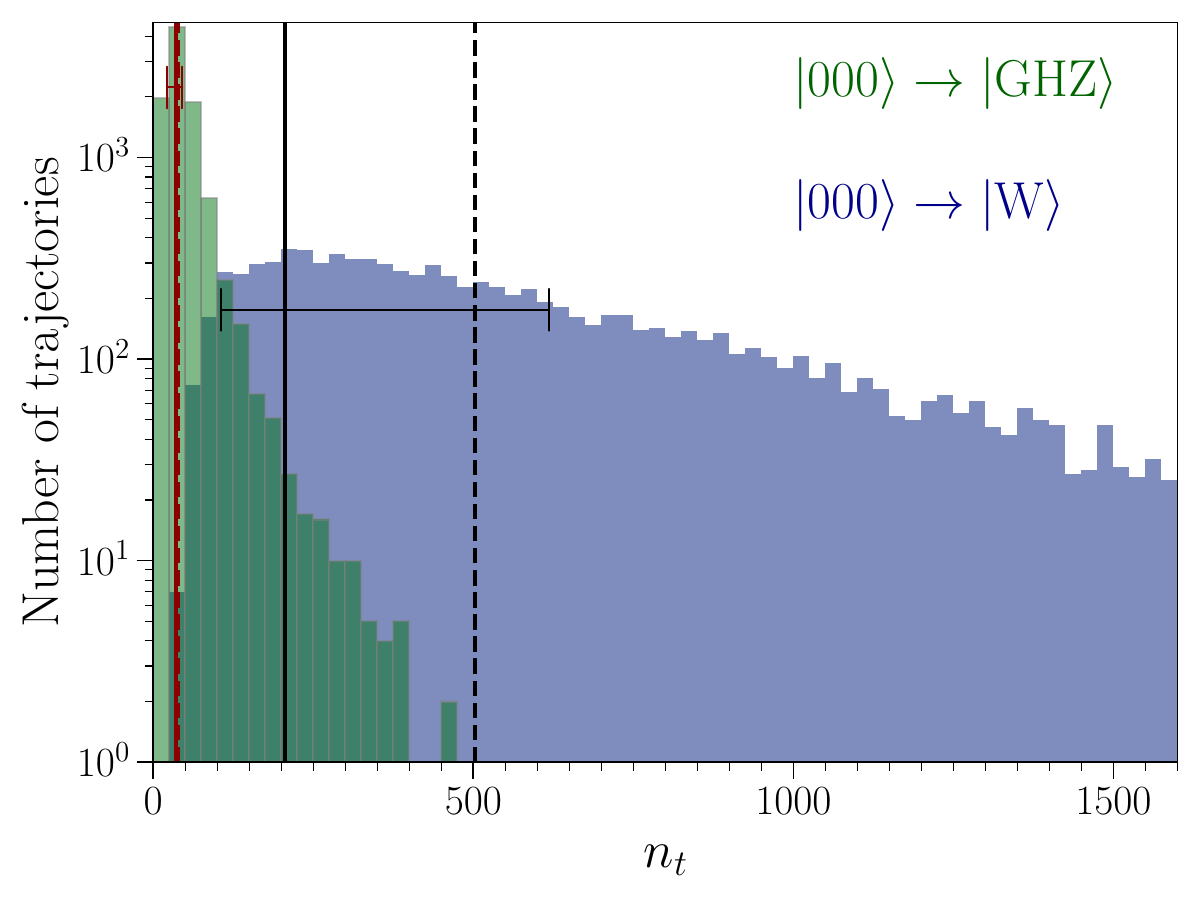}
\caption{Histogram of the number of trajectories that have reached fidelity $F^\ast=97.5\%$ in $n_t=t/\delta t$ time steps for the $N=3$ target states $|{\rm W}\rangle$ (blue, see Fig.~\ref{fig8}) 
and $|{\rm GHZ}\rangle$ (green, see Fig.~\ref{fig9}).  Using a total of $M=10^4$ runs,
each bin shows the corresponding trajectory number accumulated  over 25 subsequent time steps.
Note the logarithmic scale for the vertical axis.
Solid vertical lines (horizontal bars) indicate the respective mode $N_m$ (half-width $\Delta N$),
and dashed vertical lines show the median $N_s$. For the W state, $\beta=y$ steering operators have been 
included.}
\label{fig10}
\end{figure}

\begin{figure}
\centering
\includegraphics[width=\columnwidth]{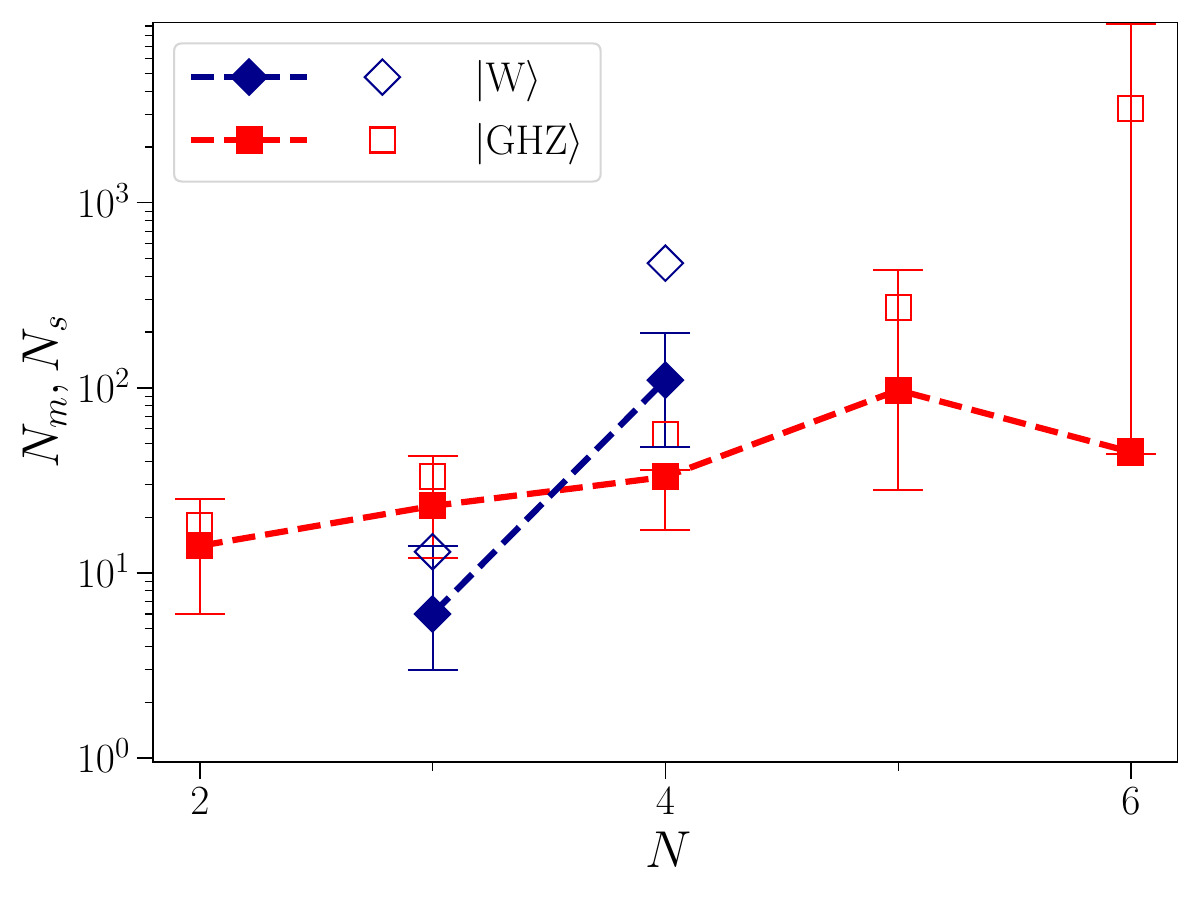}
\caption{
Mode $N_m$ (filled symbols), median $N_s$ (open symbols), and half-width (vertical bars) 
vs qubit number $N$ for active steering to the $N$-qubit states  $|{\rm W}\rangle$ and $|{\rm GHZ}\rangle$ in 
Eq.~\eqref{GHZn}. Note the logarithmic scale for the vertical axis.  For each case, the respective
target fidelity is $F^\ast=80\%$ and results have been obtained from $M=10^4$ ($M=10^3$ for $N=6$) trajectories.
Dashed lines are guides to the eye only.  The weights 
$p_r$ in Eq.~\eqref{totalcost} have been chosen as $p_{r+1}=0.1p_r$ for $1<r<N-1$,
with $p_1=0.9$ and $p_N=1-\sum_{r<N}p_r$. For the W state, 
$\beta=y$ steering operators have been included. For $N=2$,  $|{\rm Bell}\rangle$
is identical to $|{\rm GHZ}\rangle$ in Eq.~\eqref{GHZn}.  
}
\label{fig11}
\end{figure}

Results for the time evolution of the cost functions are shown in Fig.~\ref{fig8} for the W state, and in Fig.~\ref{fig9} for the GHZ state. In both cases, the fidelity threshold is $F^\ast=97.5\%$.  
For these states, the dependence of the mode $N_m$, of the median $N_s$, and of the 
half-width $\Delta N$ on the target fidelity $F^\ast$ are shown
in Fig.~\ref{fig7}.  Figure \ref{fig7} indicates that reaching the
$|{\rm W}\rangle$ state requires more steps than for the $|{\rm GHZ}\rangle$ state, 
especially when asking for high target fidelity $F^\ast$.  This observation is consistent with the fact that almost all $N=3$ states with genuine tripartite entanglement are  LOCC-related to $|{\rm GHZ}\rangle$ \cite{Dur2000}.  Preparing the much more elusive $|{\rm W}\rangle$ state thus is expected to 
be a challenging task.  By comparing the results in Figures~\ref{fig8} and \ref{fig9}, we observe
that the individual trajectories are of different character for both target states.  For the GHZ state in 
Fig.~\ref{fig9}, we find similar trajectories as for the Bell state in Fig.~\ref{fig3}: 
the trajectory cycles through highly entangled intermediate states different from the target state, but
then one quantum jump is enough to reach the final target state.  For the W state, on the other hand,
Fig.~\ref{fig8} shows that the fluctuating trajectories are qualitatively different. 
Typically, jumps to the almost perfect target state (as found for the GHZ state) are much less likely.  

\begin{figure}
\centering
\includegraphics[width=\columnwidth]{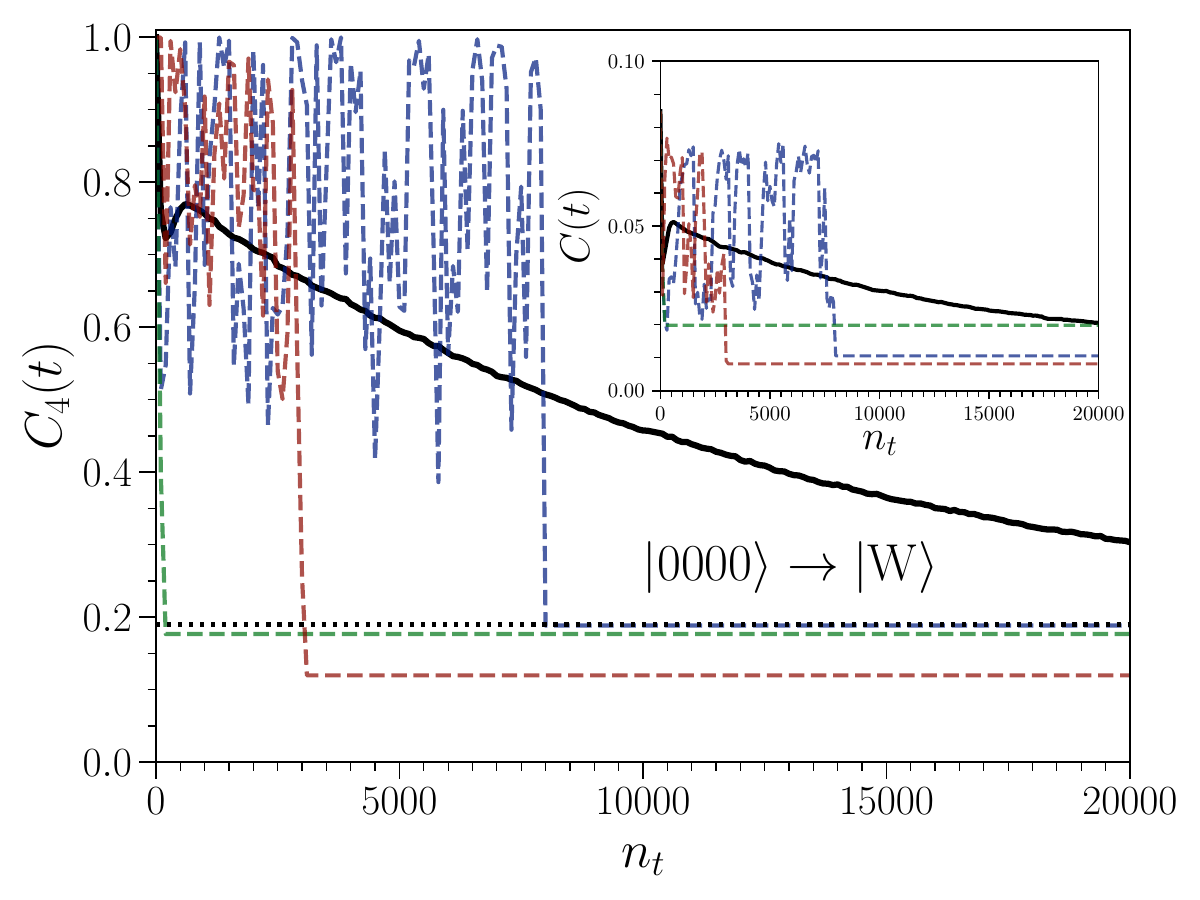}
\caption{ Active steering protocol for $N=4$ qubits and target state $|\Psi_f\rangle=|{\rm W}\rangle$ in Eq.~\eqref{GHZn} with target fidelity $F^\ast=90\%$, using the total
cost function \eqref{totalcost} with $p_1=0.9$, $p_2=0.09$, $p_3=0.009$, and $p_4=1-p_1-p_2-p_3$.
We included $\beta=y$ steering operators. 
Main panel: Global fidelity cost function $C_4(t)=1-F^2(t)$ vs number of time steps $n_t=t/\delta t$.  The dashed colored curves show three measurement-resolved  
trajectories, the solid black curve is an average over $M=10^4$ runs. 
The dotted horizontal line corresponds to the fidelity threshold.
Inset: Corresponding results for the total cost function $C(t)$.
}
\label{fig12}
\end{figure}

\begin{figure}
\centering
\includegraphics[width=\columnwidth]{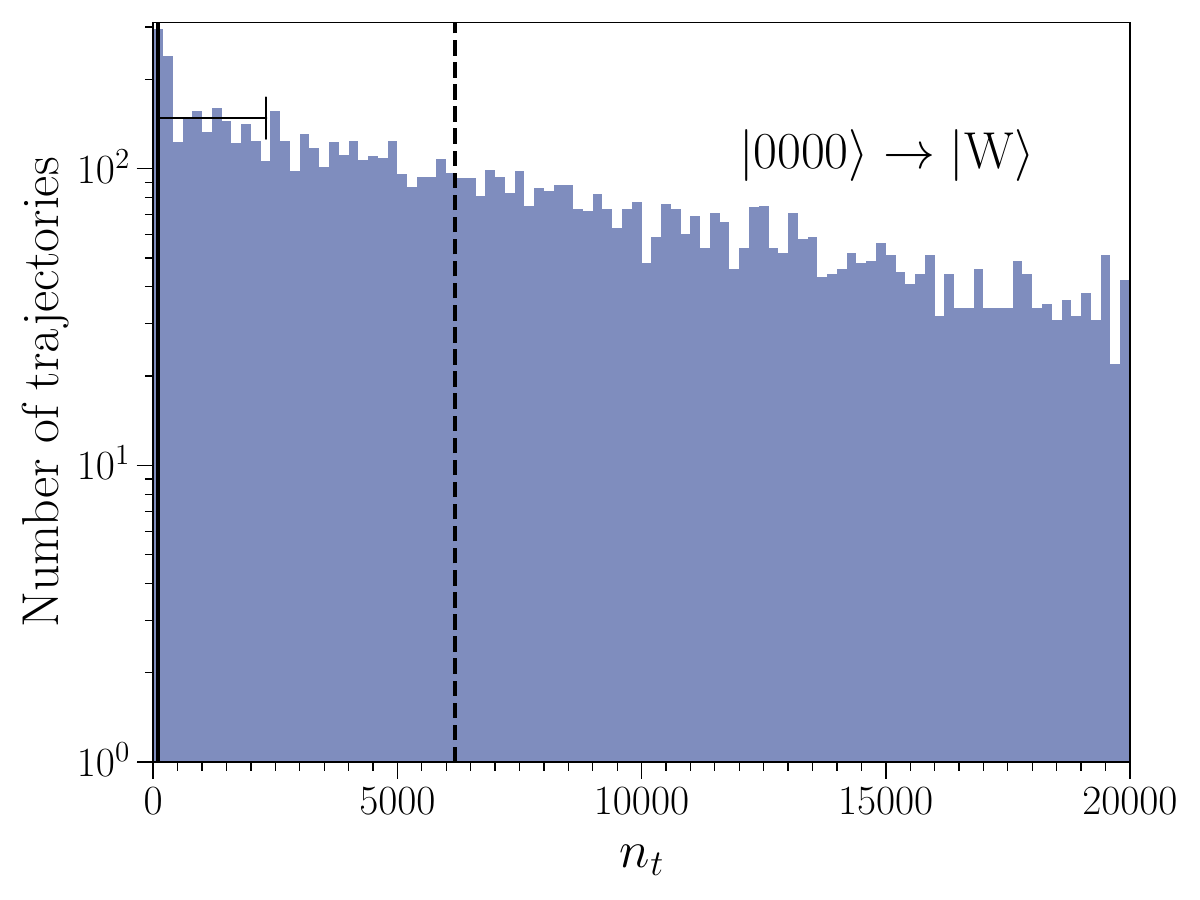}
\caption{ 
Histogram of the number of trajectories that have 
reached fidelity $F^\ast=90\%$ in $n_t=t/\delta t$ time steps from $M=10^4$ runs for
the $N=4$ target state $|{\rm W}\rangle$, see Fig.~\ref{fig12}.
Each bin shows the trajectory number accumulated over 
200 subsequent time steps, with a logarithmic scale for the vertical axis.  The solid vertical line (horizontal bar) indicates the mode $N_m$ (half-width $\Delta N$),
the dashed vertical line shows the median $N_s$.
}
\label{fig13}
\end{figure}

\begin{figure}
\centering
\includegraphics[width=\columnwidth]{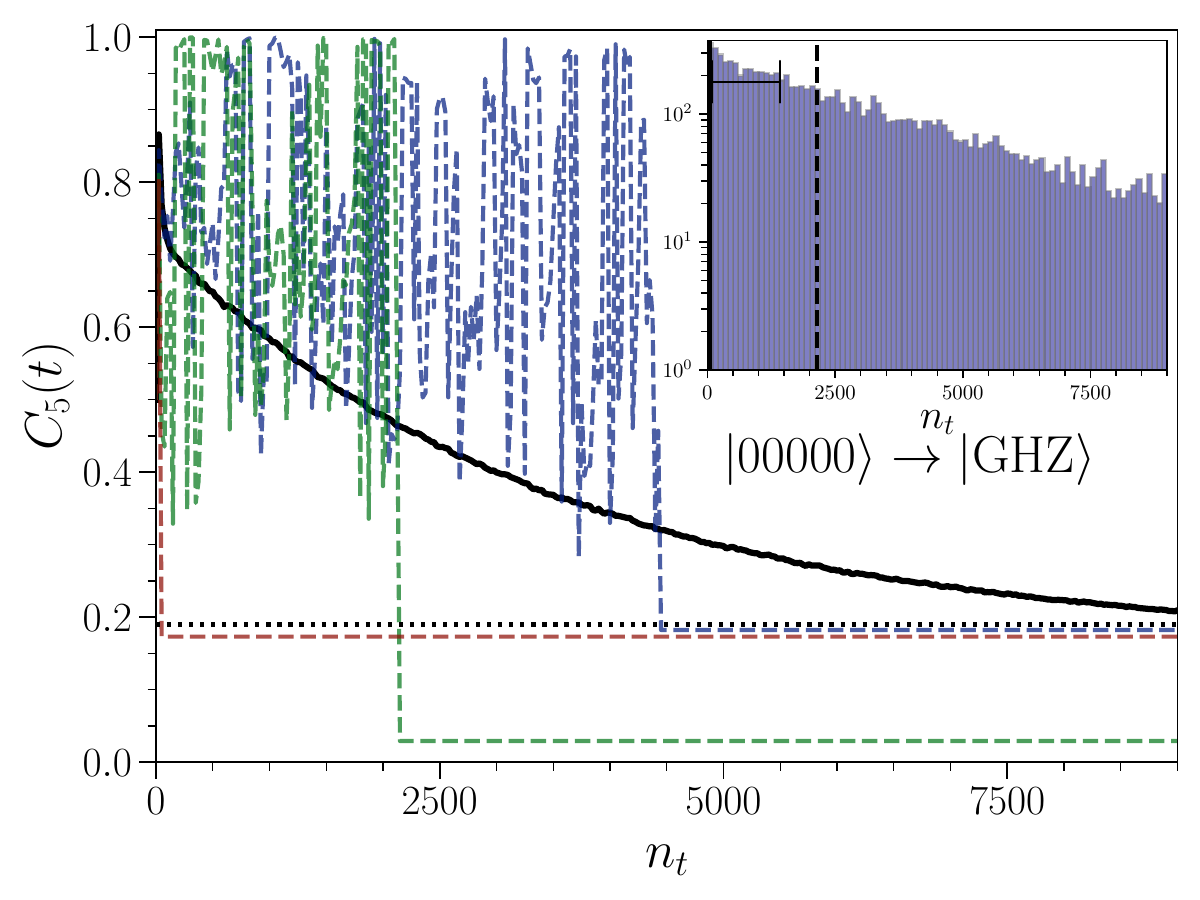}
\caption{ Active steering protocol for $N=5$ qubits and target state $|\Psi_f\rangle=|{\rm GHZ}\rangle$ in Eq.~\eqref{GHZn} with target fidelity $F^\ast=90\%$, using the total
cost function \eqref{totalcost} with $p_1=0.9$, $p_2=0.09$, $p_3=0.009$, $p_4=0.0009$, and 
$p_5=1-p_1-p_2-p_3-p_4$.
Main panel: Global fidelity cost function $C_5(t)=1-F^2(t)$ vs number of time steps $n_t=t/\delta t$.  The dashed colored curves show three measurement-resolved  
trajectories, the solid black curve is an average over $M=10^4$ runs. 
The dotted horizontal line corresponds to the fidelity threshold.
Inset: Corresponding histogram of the number of time steps needed for reaching convergence. 
Each bin shows the trajectory number accumulated over 
100 subsequent time steps, with a logarithmic scale for the vertical axis.
The solid vertical line (horizontal bar) indicates the mode $N_m$ (half-width $\Delta N$),
the dashed vertical line shows the median $N_s$.
}
\label{fig14}
\end{figure}

This difference is also manifest in a different dependence of $N_m$ and $N_s$ on the target fidelity $F^\ast$ as shown in Fig.~\ref{fig7}.  While the dependence of $N_m$ and $N_s$ on $F^\ast$ is exponential for both states,  
the rate governing the growth of $N_m(F^\ast)$ and $N_s(F^\ast)$ is very
small for the GHZ state but of significant magnitude for the W state.  Similarly, for the target fidelity $F^\ast=97.5\%$, 
the histograms in Fig.~\ref{fig10} yield $N_m= 206$ for the W state (blue histogram) but $N_m=35$
for the GHZ state (green histogram).  The GHZ state is therefore much easier to realize.   
The distribution functions are again found to be broad, asymmetric, with a non-Gaussian shape as observed for the Bell state in Fig.~\ref{fig4}.

We next apply our active steering protocol to systems with qubit number $N>3$. The weights 
$p_r$ in Eq.~\eqref{totalcost} have empirically been chosen as $p_{r+1}=0.1p_r$ for $1<r<N-1$,
with $p_1=0.9$ and $p_N=1-\sum_{r<N}p_r$, but the protocol could probably be made  
more efficient by a smarter choice for these weights.
Since the computational effort for numerical simulations 
using the present formulation of the protocol scales exponentially in $N$, 
we here limit ourselves to $N\le 6$ qubits and moderate target fidelities $F^\ast$.  We refer to Sec.~\ref{sec4B} for a discussion of variants of our protocol which should be able to reach large values of $N$.  
First, in Fig.~\ref{fig11}, choosing a rather poor but fixed fidelity threshold $F^\ast=80\%$, we 
explore how $N_m$, $N_s$, and $\Delta N$ 
 scale with the qubit number $N$ for the highly entangled $N$-qubit states $|{\rm GHZ}\rangle$ and $|{\rm W}\rangle$ in Eq.~\eqref{GHZn}.   We recall that, on general grounds,
such states cannot be prepared by physically realizable passive steering protocols \cite{Ticozzi2012}, see Sec.~\ref{sec1}.  Moreover, while the $N$-qubit GHZ state could be prepared in a simpler way by projective measurements of suitable stabilizer operators, such a route is not available for the non-stabilizer W state. 
Figure \ref{fig11} shows that upon increasing $N$ with fixed fidelity threshold, 
the values of $N_m$ and $N_s$ also increase.  Similar to our observations in 
Fig.~\ref{fig7} for the $F^\ast$-dependence of these numbers at fixed $N$, the requirements 
needed for preparing the W state are more demanding than for the GHZ state.

In Fig.~\ref{fig12}, we next show the steering dynamics for the $N=4$ W state, again
for selected measurement-resolved trajectories and for the averaged cost functions.
The corresponding histogram is shown in Fig.~\ref{fig13}.  On a qualitative level,
the steering protocol gives similar results as for the $N=3$ W state in Figs.~\ref{fig8}
and \ref{fig10}. However, even for
the modest fidelity threshold $F^\ast=90\%$ used in Figs.~\ref{fig12} and \ref{fig13},
the mode $N_m$ and median $N_s$ of the distribution are about one order of magnitude larger than
what we found for the $N=3$ W state with $F^\ast=97.5\%$, see
Figs.~\ref{fig8} and \ref{fig10}.  Finally, in Fig.~\ref{fig14}, we show
the steering dynamics and the corresponding histogram for the $N=5$ GHZ state.
The typical trajectories again feature jump-like steps as for $N=3$, see Figs.~\ref{fig9}
and \ref{fig10}.


\section{Discussion and outlook}\label{sec4}

In this section, we summarize our active steering protocol and discuss 
open points that, in our opinion, deserve to be studied by future work.  
We begin in Sec.~\ref{sec4a} with several comments on implementation aspects.  
We then continue in Sec.~\ref{sec4B} with a discussion of the scaling of the protocol
with the number $N$ of system qubits.
A summary of our key results can be found in Sec.~\ref{sec4c} 
along with perspectives for future research.

\subsection{Implementation aspects}
\label{sec4a}

While the active steering protocol laid out in Sec.~\ref{sec2} is formulated 
in a platform-independent way, experimental success will depend on  
the concrete circuit realization.  
We here discuss several salient points of applied importance for our approach.  

First, a key element of our protocol is the ability to efficiently perform
Bell measurements of detector qubit pairs. 
In many platforms, these measurements can be routinely performed by measuring the eigenvalues (syndromes) 
${\cal O}^{x,z}=\pm 1$ of the detector two-qubit Pauli operators in Eq.~\eqref{POp}.
However, it is still a challenge to perform Bell measurements in some platforms, e.g.,
in photonic circuits \cite{Nitsche2018,Pirandola2018}.  On the other hand, it is comparatively easy to provide detector qubits in the desired initial state for the photonic
platform.  We note that one may in general use different physical detector qubits during
different cycles of the protocol.

It is also important to keep in mind that these projective measurements take 
a finite time $\tau_{\rm meas}$. In fact, ideal projective measurements (with $\tau_{\rm meas}\to 0$)
come with infinite resource costs \cite{Guryanova2020}. 
 Unless a fault-tolerant platform is available, $\tau_{\rm meas}$ should be short 
 compared to the  decoherence time of the circuit, as well as to the time scales characterizing the intra-system Hamiltonian evolution.
Similar remarks apply to the initial preparation of the detector state $|00\rangle_d$
before every steering cycle.  Fast readout and qubit reset techniques (with time scales of order 100 ns) have been reported for superconducting qubit platforms \cite{Magnard2018,Sivak2023}.
In addition, for a given physical setup, one may also need to take into account the finite time needed for ramping up (and switching off) the steering operators. 
However, in a related recent experimental work using superconducting qubits \cite{Koh2023}, the respective time scales were found to be short compared to $\tau_{\rm meas}$.

Second, let us discuss the basic time scale $\delta t$ for one iteration cycle of
the protocol. On the one hand, $\delta t$ should not be too short since otherwise the 
protocol becomes inefficient. On the other hand, $\delta t$ should be small enough to 
validate the weak measurement limit and the small-$\delta t$ expansion of the Kraus operators, see 
Sec.~\ref{sec2a}.   
Choosing the gate couplings of similar magnitude, $J_n\approx J$, 
we expect that $J\delta t\sim 0.2$ should work well, cf.~Sec.~\ref{sec3}.

Third, apart from the time scales $\tau_{\rm meas}$ and $\delta t$, one also needs to 
account for the finite time $\tau_{\rm calc}$ required for the classical 
calculation performed in parallel to the experimental protocol.  
In these calculations, $\overline{dC(K)}$ is computed for all steering parameter configuations
in order to determine the best steering parameter $K$,
and the state $|\Psi(t)\rangle$ is updated after each measurement.
For the values of $N$ studied in Sec.~\ref{sec3}, we find that $\tau_{\rm calc}$
is much shorter than the expected values for $\tau_{\rm meas}$.  The 
classical computation time is thus not expected to impose restrictions.
 
Fourth, the probably most important practical restriction for our protocol at present
comes from our assumption of having a platform free from external noise and/or  static errors \cite{Edd2023}.  
In the near future, error-corrected circuits harboring several (say, a dozen) fault-tolerant
logical qubits are expected to come into reach \cite{Andersen2020,Stricker2020,Egan2021,Ryan2021,Krinner2022,Abobeih2022,Zhao2022a,Google2022,Sivak2023,Ni2023}.  Such platforms will represent an ideal playground for our protocols.
Without quantum error correction, almost noise-free platforms are available 
in trapped ion systems \cite{Monroe2021}.  In addition, novel types of noise-protected 
superconducting qubits \cite{Gyenis2021a,Gyenis2021b} may soon realize circuits with strongly reduced noise levels. Similarly, once Majorana qubits become available, 
very low noise levels are expected due to topological protection mechanisms \cite{Plugge2017,Karzig2017,Gau2020a,Gau2020b}.  

Finally, for any implementation, one should verify the ultimate success of the 
protocol. Such a benchmarking could be performed by using quantum 
state tomography methods such as shadow tomography 
\cite{Huang2020,Kliesch2021,Nguyen2022}.

\subsection{Scalability of active decision protocols}\label{sec4B}

In Sec.~\ref{sec3}, we show explicit numerical simulation results for relatively small qubit number $N\le 6$ only. Nonetheless, these results constitute a proof of concept that active steering protocols allow for the engineering of passively unsteerable states. 
Moreover, already for moderate $N\alt 10$, the preparation of exotic highly
entangled states is a very nontrivial task for which active steering provides a fresh perspective. In fact, current experimental efforts on state engineering 
typically consider quantum hardware with up to ten entangled qubits (rather than 
large-scale systems), which suffices for typical applications in quantum communication, 
quantum sensing, or quantum foundational experiments.  
Our proposal can be readily realized for such systems.  

As remarked before, it is not feasible to exactly represent and numerically simulate
the time-evolving quantum state trajectories in our active steering protocol 
on a classical computer for large-scale systems. 
Since the numerical demands grow exponentially with increasing $N$, further advances are needed to improve the performance of the protocol for significantly larger $N$.  
For instance, with an optimized choice for the probabilities $p_r$ in Eq.~\eqref{totalcost},
one may be able to significantly speed up the protocol. 
Moreover, with suitable modifications, our active feedback strategies could be used for simpler
quantum tasks than state preparation. Important examples include adiabatic and non-adiabatic state manipulation, and the realization of entanglement transitions in actively monitored circuits.
For such tasks, the protocol becomes much simpler and may enable a study of the large-$N$ limit.

One of the possible routes to scalability of the protocol is multi-block engineering:
By employing a variant of our protocol, inspired by Ref.~\cite{Smith2023}, one may consider the local steering of 
small parts of the system, where one subsequently builds up the full system state by merging different small-$N$ blocks.
Since our approach is efficient in actively steering small-$N$ blocks, 
one may be able to scale to larger system sizes in this manner.  
While this idea is speculative at present, it could eventually
obviate the need for expensive classical simulations of a large quantum system.

Furthermore, we expect that the cost functions used in the active decision policy
need not be computed exactly in order to achieve state preparation with high fidelity. 
Indeed, the cost functions in our approach provide (i) an overall ``driving bias'' for the steering landscape and (ii) a local ``potential'' to avoid trapped states. Both these main ingredients can, in principle, be simulated approximately. As a consequence, numerical methods for approximately solving quantum problems on classical hardware can be applied, e.g., tensor-network approaches, quantum Monte Carlo simulations, machine learning, or shadow tomography~\cite{Huang2020,Kliesch2021}. Such methods often remain efficient
for large-scale systems. Adapting them to our protocols and studying the resilience to those approximations are interesting topics for future research.  \\

\subsection{Summary and perspectives}\label{sec4c}

We have proposed and tested 
active steering protocols for quantum state preparation in 
monitored quantum circuits.  In our approach, entanglement
is generated in the time-evolving system state $|\Psi(t)\rangle$ by 
a weak-measurement version of entanglement swapping using
Bell measurements of detector qubit pairs. 
Active decision making strategies then determine the optimal
steering operators applied in the next time step.  
We find that in practice, a limited and realizable set of simple Pauli steering operators 
is sufficient to warrant convergence of the active steering protocol.  
We emphasize that we have used a rather restrictive notion of locality in our formulation of
the active steering protocol since only Bell measurements between adjacent detector qubits are allowed.  Interesting extensions of our protocols could permit a larger spatial coupling range.

A key finding of our analysis is that the standard fidelity is not a useful cost function for active steering to many-body
target states.  A detailed discussion of this issue can be found in Sec.~\ref{sec2c}, where we show that a useful cost function
must be able to monitor local substructures of the many-body state.  In many cases, those local structures can be efficiently 
diagnosed by monitoring weak values (which are relatively easy to compute).  In other cases, the failure of the fidelity cost
function is reminiscent of Anderson's orthogonality catastrophe.  We have demonstrated
that active steering is possible in practice when including \textit{local fidelity terms} which monitor
the $r$-body reduced density matrices (built from the time-evolving state) in comparison to the 
corresponding reduced density matrices for the target state.  In our present formulation, we constrain all possible local fidelity terms with  $1\le r\le N$.

Our numerical simulations reveal that the distribution of the number of time steps $n_t$ needed to reach convergence towards a 
desired target state is typically far from a Gaussian distribution. (Here distribution refers to an average taken over many different measurement-resolved trajectories.)
The distribution instead is found to be skewed, with a maximum (mode) $N_m$ reached after an only moderate number of steps for small $N$.  
We have also characterized the distribution by the median $N_s$ and the half-width $\Delta N$; for a 
detailed discussion, see Sec.~\ref{sec3}.

Let us then conclude by describing some perspectives for future research.  

\emph{State manifolds.---} Can active steering protocols converge toward 
state manifolds instead of a single target state? For example, such a protocol  can utilize a cost function
maximizing the entanglement entropy \eqref{EE} or other entanglement
measures instead of $C(t)$ in Eq.~\eqref{totalcost}.
The active steering protocol is then expected to target a manifold of maximally entangled
states. An interesting open question is to what extent such state manifolds resemble the ``dark manifolds'' obtained from passive steering \cite{Zanardi2014}, 
where the dark space is a degenerate Lindbladian subspace with eigenvalue zero and the final state depends on the initial state $|\Psi(0)\rangle$. For the active steering case, a different scenario may be that 
the protocol will continue cycling through all reachable 
states in the manifold as long as it is evolving in time, independent of the initial state.

We expect that state manifolds may already appear for the cost functions used above. 
Let us describe three scenarios how this can happen: 
(i) One may reduce the set of possible steering operators or the number of measurements. 
In such cases,
it is generally not possible anymore to steer towards an arbitrary target state,
and the protocol is likely to reach a state manifold.  Because of the drastic 
simplifications offered by this route, it should also be possible to tackle larger 
systems.
(ii) A similar situation arises if one omits some cost function terms in Eq.~\eqref{totalcost}.  
For instance, by omitting all local fidelity terms with $r<N$, 
one reaches the trapped-state manifold discussed in Sec.~\ref{sec2c}.
(iii) One may simultaneously steer towards two non-orthogonal target states $|\Psi_f^{(1)}\rangle$
and $|\Psi_f^{(2)}\rangle$ by adding the respective cost functions weighted with 
probabilities $w_1$ and $w_2=1-w_1$.  In that case, by varying $w_1$, 
we expect that one obtains a manifold of states interpolating between both target states. 
Since one can systematically study effects of mismatch in the target state,
such a protocol could also give useful insights about the impact of errors \cite{Edd2023}. 

\emph{State purification.---} Within our protocol, if one starts with a pure initial state, the time-evolving state remains pure at all times.   With suitable modifications, our protocol could  be used for 
state purification.  For instance, one may start from a maximally mixed (infinite temperature) initial state and use the active steering protocol to steer towards an arbitrary, possibly pure, target state.  In order to 
implement this program, one needs to adapt the present formulation of the protocol in order to allow for mixed initial states.

\emph{Multipartite entanglement.---} One can monitor the increase of  multipartite entanglement measures \cite{Nielsen2000} during the  protocol.  Entanglement structures can be built up very quickly in active steering
protocols of the type considered here since 
limitations imposed by the Lieb-Robinson bound for unitary evolution are absent, see also Ref.~\cite{Sang2023}.  
In the present work, we have only
studied the entanglement entropy, and we have only considered it for the simplest case of $N=2$ system qubits.

\emph{Memory effects.---} Another interesting generalization is to study active steering protocols where instead of the re-initialization of the detector qubit pair in the state $|00\rangle_d$ in each step,
one simply takes the previously measured Bell state as new initial state.  This scheme is
simpler to  implement since one avoids the reset of the detector qubits.  However, it also introduces memory effects for the state dynamics \cite{Doggen2023}. At present it is unclear how 
this change will affect convergence properties.

\emph{Geometrically local fidelity cost functions.---}
In the present version of the active steering protocol, we take into account all ordered subsets ${\cal M}$ of 
$r<N$ qubits when computing the local cost function $C_r(t)$ in Eq.~\eqref{localcost}.  If one truncates 
the sum over ${\cal M}$ in Eq.~\eqref{localcost} to include only geometrically local sets of nearby qubits,
significant simplifications are possible and one is able to study larger system sizes, i.e., larger $N$.
Alternatively, it is also worthwhile to explore whether it suffices to include just the $r=1$ and $r=2$
cost function terms on top of the global fidelity term.

\emph{Connections to machine learning.---}
It may be feasible to significantly improve our active steering protocol by 
 employing quantum machine learning methods \cite{Biamonte2017,Carrasquilla2021,Neupert2022}. 
Let us briefly speculate about different possibilities.
First, machine learning may be useful for optimizing the
probability weights $p_r$ appearing in the total cost 
function \eqref{totalcost}. In the present work, we have simply chosen empirical values
for these weights.
Second, instead of the exact classical computation of quantum states and the associated 
cost functions in each step (as done here), machine learning could
offer alternative strategies where such calculations are replaced 
by faster yet still accurate schemes.  
Third, in our approach, we start with a known state. 
In principle, machine learning may be employed in active protocols where 
the information on the initial state is incomplete. 
Feeding a neural network with sequences of consecutive readouts, one can perform a 
partial tomography ``on the fly" such that the missing information is reconstructed
 and the proper cost function can be estimated.

\emph{From weak to strong measurements.---}
Throughout the present work, we have assumed the weak measurement limit, 
where Kraus operators can be Taylor expanded in $J_n\delta t\ll 1$.  
On the other hand, for strong system-detector couplings $J_n$ and/or long time steps $\delta t$ such that $J_n\delta t\sim 1$, one can realize the limit of strong (projective) 
measurements \cite{Nielsen2000,Wiseman2010}.   
Understanding the crossover between the weak and strong measurement regimes 
in the context of active steering protocols raises an interesting topic for 
future research. Since the stochastic Schr\"odinger equation
\eqref{SSE} does not apply anymore outside the weak measurement limit, 
one needs to resort to Eq.~\eqref{Krausdef} without using the small-$\delta t$ form
of Kraus operators.  As suggested by the results of Ref.~\cite{Puente2024}, one may  accelerate the convergence of our active steering 
protocol by allowing for larger values of $\delta t$.  However, we leave this extension to future work.

\emph{Noise-resilient schemes.---}
Ideally, one would like to have active steering protocols that can also tolerate the 
presence of external noise without requiring a fault-tolerant platform.
In the present cost-function based approach, it is difficult
to accomodate such effects since state tracking along the time-evolving trajectory is assumed.
A conceptually different approach is to target suitable measurement operators and base the
active decision making strategy directly on measurement outcomes instead of the evaluation
of fidelity cost functions. Such a strategy would be capable of overcoming noise effects 
since the feedback now directly relies on physically measurable observables.
In addition, this approach can in principle allow for studying systems 
in the large-$N$ limit. At present, it is unclear whether such strategies 
are useful for quantum state preparation,
but they certainly can be expected to find other interesting applications.

We note that all data underlying the figures presented in this work can be retrieved at the zenodo
website \textcolor{blue}{https://zenodo.org/records/10605186}.

\begin{acknowledgments} 
 We thank Dagmar Bru{\ss}, Yaroslav Herasymenko, and Hermann Kampermann for discussions.
We acknowledge funding by the Deutsche Forschungsgemeinschaft (DFG,
German Research Foundation) under Grant No.~ 277101999, TRR 183 (project
C01), under Germany's Excellence Strategy - Cluster of Excellence Matter
and Light for Quantum Computing (ML4Q) EXC 2004/1 - 390534769, and under Grants No.~EG 96/13-1,  No.~GO 1405/6-1, and No.~RO 2247/11-1. In addition, we acknowledge funding by the Israel Science Foundation 
and by the  NSF-BSF  grant DMR-2338819.  
\end{acknowledgments}

\appendix

\section{On entanglement swapping}\label{app0}

We here give a detailed picture for how entanglement is built up in the system state $|\Psi(t)\rangle$ during our entanglement swapping protocol.
In Fig.~\ref{fig2}(a)-(c), we show the first cycle of the protocol for the case $N=2$, starting at the initial time $t=0$. 
(a) One starts from the simple product state 
\begin{align} \nonumber
|\psi(0)\rangle&=|\Psi(0)\rangle_s  \otimes |\Phi(0)\rangle_d \\
& \nonumber = \left(|0_1\rangle_s \otimes |0_2\rangle_s\right)
\otimes \left(|0_1\rangle_d \otimes |0_2\rangle_d\right)
\\ &= \left(|0_1\rangle_s\otimes |0_1\rangle_d\right) \otimes \left(|0_2\rangle_s\otimes |0_2\rangle_d\right),
\label{psia}
\end{align}
where $|\Psi\rangle_s$ and $|\Phi\rangle_d$ describe the system and detector qubits, respectively.
(In this Appendix, we use the subscript $s$ to explicitly mark system states.) 
In the last step, states are grouped to form system-detector pairs. 
(b) Switching on the selected steering operators during a time step of duration $\delta t$, each system-detector qubit pair is 
entangled by the unitary time evolution.  One obtains a product state of two entangled system-detector pairs, 
$|\psi(\delta t)\rangle=|\psi_1(\delta t)\rangle \otimes |\psi_2(\delta t)\rangle$,
where the state of pair $i=1,2$ can be written as
\begin{align}
    |\psi_i(\delta t)\rangle&=\mathcal{A}_i\, |0_i\rangle_s \otimes |0_i\rangle_d 
    + \mathcal{B}_i\, |0_i\rangle_s \otimes |1_i\rangle_d\notag \\
    &+\ \mathcal{C}_i\, |1_i\rangle_s \otimes |0_i\rangle_d
    + \mathcal{D}_i\, |1_i\rangle_s \otimes |1_i\rangle_d.
    \label{psi-i}
\end{align}
The amplitudes satisfy the normalization condition 
$|\mathcal{A}_i|^2+| \mathcal{B}_i|^2+|\mathcal{C}_i|^2+|\mathcal{D}_i|^2=1$ and are determined by the chosen steering operators.
Entanglement between the system-detector qubit pair $i$ 
means that $|\psi_i(\delta t)\rangle$ in Eq.~(\ref{psi-i}) does not factorize into a product state, i.e., 
$|\psi_i\rangle\neq |\psi_{s,i}\rangle  \otimes |\psi_{d,i}\rangle$.
The state $|\psi(\delta t)\rangle$ can equivalently be represented in the Bell basis $|\Phi_{\xi,\eta}\rangle_d$ of the detectors, see Eq.~\eqref{bell}. We thereby obtain
\begin{equation}
    |\psi(\delta t)\rangle= \sum_{\xi=0,1} \sum_{\eta=\pm 1} |\Psi^{(\xi,\eta)}\rangle_s\otimes|\Phi_{\xi,\eta}\rangle_d,
    \label{psi-Bell-1}
\end{equation}
where the ``coefficients'' of the expansion are states of the two system qubits. We  label these by superscripts, $|\Psi^{(\xi,\eta)}\rangle_s$, in order to distinguish them from 
Bell states  of the system qubits, $|\Psi_{\xi,\eta}\rangle_s$, defined in  analogy to Eq.~\eqref{bell}.  Explicitly, from Eq.~\eqref{psi-i}, we find
\begin{widetext}
\begin{eqnarray*}
   |\Psi^{(0,+)}\rangle_s &=& \frac{1}{\sqrt{2}} 
\Bigl[ \left(\mathcal{A}_1\mathcal{A}_2+\mathcal{B}_1\mathcal{B}_2 \right)|0_1\rangle_s \otimes |0_2\rangle_s 
+ \left(\mathcal{A}_1\mathcal{C}_2+\mathcal{B}_1\mathcal{D}_2 \right)|0_1\rangle_s \otimes |1_2\rangle_s+\\
&&\quad + \left(\mathcal{C}_1\mathcal{A}_2+\mathcal{D}_1\mathcal{B}_2 \right)|1_1\rangle_s \otimes |0_2\rangle_s + \left(\mathcal{C}_1\mathcal{C}_2+\mathcal{D}_1\mathcal{D}_2 \right)|1_1\rangle_s \otimes |1_2\rangle_s
   \Bigr],
   \\
   |\Psi^{(0,-)}\rangle_s&=& \frac{1}{\sqrt{2}}
\Bigl[\left(\mathcal{A}_1\mathcal{A}_2-\mathcal{B}_1\mathcal{B}_2 \right)|0_1\rangle_s \otimes |0_2\rangle_s + \left(\mathcal{A}_1\mathcal{C}_2-\mathcal{B}_1\mathcal{D}_2 \right)|0_1\rangle_s \otimes |1_2\rangle_s+\\ && \quad +\left(\mathcal{C}_1\mathcal{A}_2-\mathcal{D}_1\mathcal{B}_2 \right)|1_1\rangle_s \otimes |0_2\rangle_s + \left(\mathcal{C}_1\mathcal{C}_2-\mathcal{D}_1\mathcal{D}_2 \right)|1_1\rangle_s \otimes |1_2\rangle_s
   \Bigr],
   \\
    |\Psi^{(1,+)}\rangle_s &=& \frac{1}{\sqrt{2}}
\Bigl[\left(\mathcal{A}_1\mathcal{B}_2+\mathcal{B}_1\mathcal{A}_2 \right)|0_1\rangle_s \otimes |0_2\rangle_s + \left(\mathcal{A}_1\mathcal{D}_2+\mathcal{B}_1\mathcal{C}_2 \right)|0_1\rangle_s \otimes |1_2\rangle_s+\\ &&\quad 
+\left(\mathcal{C}_1\mathcal{B}_2+\mathcal{D}_1\mathcal{A}_2 \right)|1_1\rangle_s \otimes |0_2\rangle_s + \left(\mathcal{C}_1\mathcal{D}_2+\mathcal{D}_1\mathcal{C}_2 \right)|1_1\rangle_s \otimes |1_2\rangle_s \Bigr],
   \\
   |\Psi^{(1,-)}\rangle_s &=& \frac{1}{\sqrt{2}}
\Bigl[\left(\mathcal{A}_1\mathcal{B}_2-\mathcal{B}_1\mathcal{A}_2 \right)|0_1\rangle_s \otimes |0_2\rangle_s + \left(\mathcal{A}_1\mathcal{D}_2-\mathcal{B}_1\mathcal{C}_2 \right)|0_1\rangle_s \otimes |1_2\rangle_s + \\ &&\quad
+\left(\mathcal{C}_1\mathcal{B}_2-\mathcal{D}_1\mathcal{A}_2 \right)|1_1\rangle_s \otimes |0_2\rangle_s + \left(\mathcal{C}_1\mathcal{D}_2-\mathcal{D}_1\mathcal{C}_2 \right)|1_1\rangle_s \otimes |1_2\rangle_s
   \Bigr].
\end{eqnarray*}
\end{widetext}

Generically, $|\Psi^{(\xi,\eta)}\rangle_s$ describes an entangled state that can also be expanded in the Bell basis  $|\Psi_{\xi',\eta'}\rangle_s$, leading to a superposition of products $|\Psi_{\xi',\eta'}\rangle_s \otimes |\Phi_{\xi,\eta}\rangle_d$. Since in our general framework for $N>2$,  we do not employ Bell states for system qubits, 
the explicit form of this superposition is not given here.   
(c) The effect of the Bell measurement on the detector qubits at time $t=\delta t + 0^+$ is a projection of the state (\ref{psi-Bell-1}) to one of the Bell states,
depending on the measurement outcome $(\xi,\eta)$, thus removing the three other components. 
As a result, the total wave function collapses to a product state (no summation over repeated indices), 
\begin{align}
 |\psi(\delta t)\rangle \rightarrow  \  |\psi(\delta t+0^+)\rangle=\mathcal{N}_{\xi,\eta}|\Psi^{(\xi,\eta)}\rangle_s\otimes|\Phi_{\xi,\eta}\rangle_d,
\end{align}
where $\mathcal{N}_{\xi,\eta}$ is a normalization factor.
The system qubits are now in the entangled state $\mathcal{N}_{\xi,\eta}|\Psi^{(\xi,\eta)}\rangle_s$.

If one instead starts with an already entangled state of the two qubits, as for the second cycle at time $t=\delta t$ shown in Fig.~\ref{fig2}(d-f), the state after the joint time evolution of system-detector qubits can still be expanded in the Bell basis of detector qubit states, see Eq.~(\ref{psi-Bell-1}). 
However, the states $|\Psi^{(\xi,\eta)}\rangle_s$ will now be determined by more amplitudes, 
including those parametrizing the initial state $|\Psi(\delta t)\rangle$. Performing the Bell measurement  
then again yields a product state between the system and detector parts, where $|\Psi^{(\xi,\eta)}\rangle_s$ is determined by the measurement outcome $(\xi,\eta)$.

Finally, let us note that because of the so-called \textit{entanglement monogamy} \cite{Coffman2000, Osborne2006, Seevinck2010, Horodecki2009}, the degree of entanglement between detectors and the corresponding system qubits is reduced in panel (e) of Fig.~\ref{fig2} as compared to panel (b), since the system qubits are already entangled. 
As a result, additional entanglement generated between the system qubits after the Bell measurement of detectors is expected to be weaker than the initially generated
system qubit entanglement indicated in panel (c). 
Importantly, weak measurements of already entangled system qubits, followed by projective  Bell measurements of the corresponding detector qubits (referred to as entanglement swapping), 
do not prohibit entanglement generation in the system. \\

\section{Single-detector protocol}\label{app1}

We here briefly discuss an alternative protocol, where only a \emph{single}
detector qubit ($\tau$) is coupled to \emph{two} system qubits during a given step 
of the steering protocol. 
For clarity, let us consider system qubits $\sigma_{1}$ and $\sigma_2$ 
as the steered qubit pair. A general steering operator for this case can be written 
in the form 
\begin{equation}\label{frustratedsteering}
    H_K = J \left[ \cos (\theta) \sigma_1^{\alpha_1}\tau^{z} +
    \sin(\theta) \sigma_2^{\alpha_2} \tau^{x} \right] ,    
\end{equation}
with discrete steering parameters $(\alpha_1,\alpha_2)$ (with $\alpha=x,y,z$) 
and a continuous parameter $\theta\in [0,\pi/2]$.  The overall coupling $J$ is considered
fixed and known.  In each time step of the protocol, one then has a unitary evolution under $H_K$.
Note that in general, the two pieces in Eq.~\eqref{frustratedsteering} do not 
commute, which implies that we have a ``frustrated steering'' protocol \cite{Roy2020}. 
In the most general case, we assume that the detector qubit is
initialized before each time step in the state
\begin{equation}\label{initialancilla}
    |\Phi\rangle_d = \cos (\eta) |0\rangle_d + e^{i\psi} \sin(\eta)|1\rangle_d,
\end{equation}
with fixed and known angles $(\eta,\psi)$.  Instead of the Bell measurements, 
one here performs a projective measurement of the single-qubit
detector operator $\tilde \tau_z$ after the time step $\delta t$.
We express $\tilde \tau_z$ in a general basis, where the outcome $\xi=\pm 1$ corresponds
to finding the respective orthogonal detector state, 
\begin{eqnarray}\nonumber
   |\tilde\Phi_{+}\rangle_d &=& \cos (\tilde\eta) |0\rangle_d + e^{i\tilde\psi} \sin(\tilde\eta)|1\rangle_d, \\
   |\tilde\Phi_{-}\rangle_d &=& \sin (\tilde\eta) |0\rangle_d - e^{-i\tilde\psi} \cos(\tilde\eta)|1\rangle_d,
\end{eqnarray}
with fixed and known angles $(\tilde \eta,\tilde \psi)$.
The total system-plus-detector state after the measurement outcome $\xi$ is then given by 
$|\Psi_\xi(t+\delta t)\rangle\otimes |\tilde \Phi_\xi\rangle_d$, where the index $\xi$ 
on the system state emphasizes its outcome dependence.  

One may naively expect that it is now possible to engineer arbitrary target states 
by such an approach. However, this is not possible, even when allowing
for arbitrary $\delta t$ in Eq.~\eqref{Krausdef}.  By evaluating the Kraus operators, we
find 
\begin{equation}\label{SSE2}
    |\Psi_\xi(t+\delta t)\rangle= \sum_{n=1,2} a_\xi^{(n)} e^{-i\Lambda_\xi^{(n)} \sigma_n^{\alpha_n}}|\Psi(t)\rangle,
\end{equation}
with complex-valued  amplitudes $a_\xi^{(1,2)}$ and angles $\Lambda_\xi^{(1,2)}$.  
For instance, for $\psi=\tilde \psi=0$ and $\tilde\eta=-\eta$, we find 
\begin{eqnarray}\nonumber
    \tan\Lambda^{(1,2)}_+ &=& \frac{\cos\theta}{\cos(2\eta)} \tan(J\delta t),\\
    \quad \tan\Lambda^{(1,2)}_- &=& \frac{\sin\theta}{\sin(2\eta)} \tan(J\delta t),
\end{eqnarray}
but general expressions can be written down for arbitrary parameters $(\psi,\eta,\tilde\psi,\tilde \eta)$.

According to Eq.~\eqref{SSE2}, the system state evolves by linear transformations acting separately 
on both qubits. These transformations exhaust all possible actions induced by such a  
steering protocol for $N=2$.  However, these actions are 
not sufficient for reaching arbitrary target states from a given initial state.
Consider, e.g., the initial state $|\Psi(0)\rangle=|00\rangle$ and the target state
$|\Psi_f\rangle=|{\rm Bell}\rangle$ in Eq.~\eqref{EPR}. 
By (possibly repeated) application of the time evolution in Eq.~\eqref{SSE2}, 
the state at a later time $t$ must be of the form
\begin{equation} \label{PS}
    |\Psi(t)\rangle = \left( z_{0}^{(1)} |0\rangle_1 + z_{1}^{(1)}|1\rangle_1\right)
    \otimes  \left( z_{0}^{(2)} |0\rangle_2 + z_{1}^{(2)}|1\rangle_2\right)
\end{equation}
with complex coefficients $z_{0,1}^{(1,2)}(t)$. (Their precise value is not of interest here.) 
For steering towards $|\Psi_f\rangle=|{\rm Bell}\rangle$, we observe from Eq.~\eqref{PS} that 
the terms $\sim |01\rangle, |10\rangle$ have to be projected away since there is no possibility
to otherwise arrive at $|{\rm Bell}\rangle$.
However, such an operation is missing in Eq.~\eqref{SSE2}.
A similar argument can be given for the backward direction, $|{\rm Bell}\rangle\to |00\rangle$.  

We conclude that active steering protocols employing a single detector with the steering operator $H_K$ in Eq.~\eqref{frustratedsteering} are not able to realize arbitrary operations, not even for $N=2$ qubits.
The two-detector scheme with Bell measurements illustrated in Fig.~\ref{fig2} does not suffer from 
such restrictions.

\section{Bloch tensor changes}\label{app2}

We here provide explicit expressions for $\overline{dR_{\cal S}^2}$, where the string operator ${\cal S}$ in Eq.~\eqref{string} is parametrized in terms 
of the $N$ indices $\mu_j\in \{0,1,2,3\}$.   
The corresponding expressions for RDMs needed in Eq.~\eqref{davcr} follow by setting $\mu_j=0$ for all traced-out qubits, cf.~Sec.~\ref{sec2b}. 

Assuming that steering operators and measurements are applied to the qubit pair $(n,n+1)$, we first specify the 
measurement-conditioned change $dR_{\cal S}$  of the rank-$N$ Bloch tensor in one step, see 
Eq.~\eqref{drs}.
From Eq.~\eqref{drho2} with $\alpha,\alpha'\in\{1,2,3\}$, we find
\begin{widetext}
\begin{eqnarray}\nonumber
dR_{\cal S} &=&  -2 \sum_{m=n,n+1}  \sum_{\alpha\ne \alpha_m} 
\left( s_mJ_m\delta t\, \delta_{\beta_m,z} \sum_{\alpha'}\varepsilon_{\alpha_m\alpha\alpha'} 
+ \Gamma_m \frac{\xi}{\langle c^\dagger_\eta c_\eta^{}\rangle} \delta_{\beta_m,\perp} \right)
 \, \delta_{\mu_m,\alpha} \, R_{\cal S} \\  &+& 2\eta\sqrt{\Gamma_n\Gamma_{n+1}}  \delta_{\beta_n,x}\delta_{\beta_{n+1},x}  
 \left(  \frac{ \xi}{\langle c_\eta^\dagger c_\eta^{}\rangle }  
\left ( F_{\cal S}-Q_{n,n+1} R_{\cal S}\right )-\delta t     
\left(H_{\cal S}-Q_{n,n+1} R_{\cal S}\right ) \right),\label{dRexpl}
\end{eqnarray}
\end{widetext}
with $\langle c^\dagger_\eta c_\eta^{}\rangle$ in Eq.~\eqref{avcdc}. 
For simplicity, Eq.~\eqref{dRexpl} assumes only steering operators with $\beta_m\ne y$.  
However, the final results, Eqs.~\eqref{davR} and \eqref{squared}, are specified for the general case.
The correlator \eqref{spincor} is encoded by 
\begin{equation}
    Q_{n,n+1} = Q^{\alpha_1,\alpha_2}_{n,n+1}=R_{0,\ldots,0, \alpha_n,\alpha_{n+1},0,\ldots,0},
\end{equation}
and the rank-$N$ tensors $F_{\cal S}$ and $H_{\cal S}$ in Eq.~\eqref{dRexpl} are given by
\begin{eqnarray}\nonumber
F_{\cal S}& =& \frac{1}{2^{N+1}} \sum_{{\cal S}'} R_{{\cal S}'} {\rm Tr}\left( 
(\sigma_n^{\alpha_n} {\cal S}' \sigma_{n+1}^{\alpha_{n+1}} + \sigma_{n+1}^{\alpha_{n+1}} {\cal S}' \sigma_n^{\alpha_n} ) {\cal S}
\right) ,\\ \label{Fdef}
H_{\cal S} &=& \frac{1}{2^{N+1}} \sum_{{\cal S'}} R_{{\cal S}'}  {\rm Tr}\left( \{ \sigma_n^{\alpha_n}
    \sigma_{n+1}^{\alpha_{n+1}} , {\cal S}' \} {\cal S} \right).
\end{eqnarray}
Since only  tensor components involving the indices $\mu_n$ or $\mu_{n+1}$ are affected, we write 
$F_{\mu_n,\mu_{n+1}}$ as shorthand for $F_{\cal S}$, and similarly for $R_{\cal S}$ and $H_{\cal S}$,
that is, unchanged indices $\mu_1,\ldots,\mu_{n-1}$ and $\mu_{n+2},\ldots,\mu_N$ are kept implicit.
The non-vanishing matrix elements of $F_{\cal S}$ are given by
\begin{eqnarray}\nonumber 
F_{0,0} &=& R_{\alpha_n,\alpha_{n+1}},\quad F_{\alpha_n,\alpha_{n+1}} = R_{0,0}, \\  \label{Fmatr}
F_{\alpha_n,0} &=& R_{0,\alpha_{n+1}},\quad  F_{0,\alpha_{n+1}} = R_{\alpha_n,0},\\ \nonumber
F_{\alpha\ne\alpha_n,\alpha'\ne \alpha_{n+1}} &=& \sum_{\tilde\alpha,\tilde\alpha'} \varepsilon_{\alpha_n,\alpha,\tilde\alpha} 
\,\varepsilon_{\alpha_{n+1},\alpha',\tilde\alpha'}\, R_{\tilde\alpha,\tilde\alpha'}.
\end{eqnarray} 
Similarly, $H_{\cal S}=F_{\cal S}$ except for a sign change in the  last line of Eq.~\eqref{Fmatr},  $H_{\alpha,\alpha'}=-F_{\alpha,\alpha'}$.

Taking the average over measurement outcomes in Eq.~\eqref{dRexpl},  we   arrive at Eq.~\eqref{davR}. 
Next we compute $\overline{dR_{\cal S}^2}$ from Eq.~\eqref{dRexpl}.  Recalling that $\xi^2=\xi$, only the contribution $\propto \xi$ can generate a contribution to leading order in $\delta t$. We therefore  obtain
\begin{equation}\label{squared}
  \frac12 \overline{dR_{\cal S}^2} = \delta t  
   \sum_{\eta=\pm} \frac{   G^{(\eta) 2}_{\cal S} } {\langle c_\eta^\dagger c_\eta^{}\rangle} ,  
\end{equation}
with
\begin{eqnarray}  \nonumber
    G_{\cal S}^{(\eta=\pm)} &=&  -\sum_{m=n,n+1} \sum_{\alpha\ne \alpha_m} \Gamma_m \delta_{\beta_m,\perp} \delta_{\mu_m,\alpha}  R_{\cal S} + \\
\nonumber
&+& \eta \sqrt{\Gamma_n\Gamma_{n+1}}  ( \delta_{\beta_n,x}\delta_{\beta_{n+1},x}+
    \delta_{\beta_{n},y}\delta_{\beta_{n+1},y} )  \\
    &\times&
\left(F_{\cal S}-Q_{n,n+1} R_{\cal S}\right ) . \label{Gdef}
\end{eqnarray}
We note that the denominator in Eq.~\eqref{squared} may vanish in special cases (for instance, 
if $\beta_n=\beta_{n+1}=z$), but then the numerator will also vanish and $G_{\cal S}^{(\eta)}=0$.   Similarly, for $\Gamma_n=\Gamma_{n+1}$ and $|Q_{n,n+1}|=1$, 
one of the two ratios in Eq.~\eqref{squared} is of ``$0/0$'' type. However, the final result 
is well-defined and finite.

To give an example, let us consider the case $N=3$, with qubits $(1,2)$ being steered.  Here  a
contribution from qubit 3 to the expected single-qubit RDM cost function change can be present.  
We call this term $\overline{dC^{(3)}_1(K)}$, which is due to the subset ${\cal M}=\{3\}$ 
and comes exclusively from Eq.~\eqref{squared}. We note that Eq.~\eqref{Gdef}, together with $F_{0,0,\mu}=R_{\alpha_1,\alpha_2,\mu}$, see Eq.~\eqref{Fdef}, yields 
\begin{eqnarray*} 
 && \overline{dC_1^{(3)}(K)} =   \sum_{\mu=0}^3 \frac{\overline{dR_{0,0,\mu}^2}}{4}  =   ( \delta_{\beta_1,x}\delta_{\beta_{2},x}+
    \delta_{\beta_{1},y}\delta_{\beta_{2},y} )\, \\ \nonumber &&\times\frac{\Gamma_1\Gamma_2(\Gamma_1+\Gamma_2)\delta t}{(\Gamma_1+\Gamma_2)^2-4\Gamma_1\Gamma_2 Q_{1,2}^2}
    \sum_{\alpha} (R_{\alpha_1,\alpha_2,\alpha}-Q_{1,2} R_{0,0,\alpha})^2  
\end{eqnarray*}
where the $\mu=0$ term does not contribute because of $F_{0,0,0}-Q_{1,2} R_{0,0,0}= Q_{1,2}-Q_{1,2}=0$.
Note that $R_{0,0,\alpha}$ is the respective component of the Bloch vector for qubit 3, while $R_{\alpha_1,\alpha_2,\alpha}$ encodes the entanglement between qubit 3 and the other qubits.  We conclude that the 
expected changes $\overline{dC_r(K)}$ of local cost functions ($r<N$) 
are capable of detecting entanglement features outside the reach of fidelity-based cost functions.

\bibliography{biblio}

\end{document}